\documentclass[pre,reprint,amsmath,floatfix,superscriptaddress,nofootinbib, aps]{revtex4-1}

\usepackage{chemformula} 
\usepackage[T1]{fontenc} 
\usepackage[utf8]{inputenc}
\usepackage[]{graphicx}
\usepackage{amsmath,amsfonts,amssymb}
\begin{document}
\pdfoutput=1

\newcommand*\mycommand[1]{\texttt{\emph{#1}}}


\author{Uwe Thiele}
\email{u.thiele@uni-muenster.de}
\affiliation{Institut f\"ur Theoretische Physik, Westf\"alische Wilhelms-Universit\"at M\"unster, Wilhelm Klemm Str.\ 9, 48149 M\"unster, Germany}
\affiliation{Center of Nonlinear Science (CeNoS), Westf{\"a}lische Wilhelms-Universit\"at M\"unster, Corrensstr.\ 2, 48149 M\"unster, Germany}
\affiliation{Center for Multiscale Theory and Computation (CMTC), Westf{\"a}lische Wilhelms-Universit\"at, Corrensstr.\ 40, 48149 M\"unster, Germany}

\author{Jacco H. Snoeijer}
\affiliation{Physics of Fluids Group and J. M. Burgers Centre for Fluid Dynamics,
University of Twente, P.O. Box 217, 7500 AE Enschede, The Netherlands}

\author{Sarah Trinschek}
\affiliation{Institut f\"ur Theoretische Physik, Westf\"alische Wilhelms-Universit\"at M\"unster, Wilhelm Klemm Str.\ 9, 48149 M\"unster, Germany}
\affiliation{Universit\'e Grenoble-Alpes, CNRS, Laboratoire Interdisciplinaire de Physique, 38000 Grenoble, France}

\author{Karin John}
\affiliation{Universit\'e Grenoble-Alpes, CNRS, Laboratoire Interdisciplinaire de Physique, 38000 Grenoble, France}


\title{Equilibrium contact angle and adsorption layer properties with surfactants }

\begin{abstract}
 The three-phase contact line of a droplet on a smooth surface can be characterized by the Young-Dupr\'e equation. It relates the interfacial energies with the macroscopic contact angle $\theta_e$. On the mesoscale, wettability is
  modeled by a film-height-dependent wetting energy $f(h)$. Macro- and
  mesoscale description are consistent if
  $\gamma \cos \theta_\mathrm{e} =\gamma+f(h_\mathrm{a})$ where $\gamma$ and
  $h_\mathrm{a}$ are the liquid-gas interface energy and the thickness of the
  equilibrium liquid adsorption layer, respectively.

Here, we derive a similar consistency condition for the case of a
liquid covered by an insoluble surfactant. At equilibrium, the surfactant is spatially
inhomogeneously distributed implying a non-trivial dependence of
$\theta_\mathrm{e}$ on surfactant concentration. We derive macroscopic
and mesoscopic descriptions of a contact line at equilibrium and show
that they are only consistent if a particular dependence of the
wetting energy on the surfactant concentration is imposed.This is illustrated by a simple example of dilute surfactants, for which we show excellent agreement between theory and time-dependent numerical simulations.
%
\end{abstract}

\maketitle

\section{Introduction}
Surfactants are amphiphilic molecules or particles that adsorb at
interfaces, thereby decreasing the surface tension of the
interface. Their chemico-physical properties crucially alter the
dynamics of thin liquid films with free surfaces, a fact that is
exploited for many industrial and biomedical applications,
e.g. coating, deposition or drying processes on surfaces, surfactant
replacement therapy for premature infants (see
\cite{CrMa2009rmp,MaCr2009sm} for reviews).
However, the detailed mechanism of surfactant driven flows is still an
active field of research, experimentally and theoretically. In the
simplest case, the spreading of surfactant laden droplets on solid
surfaces, the presence of surfactants leads to deviations from the
Tanner law, i.e. the spreading rate is rather $R(t)\sim t^{(1/4)}$ instead of
$R(t)\sim t^{(1/10)}$ as expected for the pure liquid (see \cite{MaCr2009sm} for review).
The basic explanation for this phenomenon is that gradients in the
surface tension are associated with interfacial (Marangoni) stresses
which drive the fluid flow and the convective and diffusive transport
of surfactant molecules along the interface. The surfactant
concentration and the interfacial tension are related by an equation
of state. 

Besides the modified Tanner law, the interplay between surfactant
dynamics and free surface thin film flows leads to a variety of
intriguing phenomena, such as surfactant induced fingering of
spreading droplets
\cite{MaLe1981cec,TrWS1989prl,MaCr2009sm,TrHS1990prl,MaCr2009sm,CCBV1999csa,CaCa1999l},
superspreading of aqueous droplets on hydrophobic surfaces
\cite{Hill1998coc,RSBM2002l}, or autophobing of aqueous drops on
hydrophilic substrates \cite{AALM2004l,CrMa2007l,BDSE2016sm}. In
addition to creating Marangoni-stresses at the free interface, several
other properties of surfactants enrich the spectrum of dynamical
behaviors observed.  Bulk solubility, their propensity to form
micelles or lamellar structures at high concentrations, the surfactant
mobility on the solid surface and their ability to spread through the
three-phase contact region are all key parameters to influence the
flow properties.  But the presence of surfactants does not only affect
the flow dynamics. Also in the static situation of a
surfactant-covered droplet on a substrate in equilibrium, the
spatially inhomogeneous distribution of surfactant will cause a
non-trivial dependence of the contact angle on the surfactant
concentration.

The governing equations that describe film flows and surfactant
dynamics at low surfactant concentrations and in situations where the
influence of wettability is negligible are well established (see
\cite{ThAP2012pf, thap2016prf} for review).
Typically, the dynamics of the liquid with a free
surface is described using an evolution equation for the film height
(derived from the lubrication approximation of a viscous Stokes flow with no-slip boundary
condition at the substrate) coupled to an evolution equation of the
surfactant concentration. The equations usually include capillarity (with a constant
surface tension, though) and Marangoni stresses via an equation of state
for the surfactant. Some models include wettability via a disjoining
pressure \cite{WaCM2002pof,CrMa2007l,JeGr1992jfm}. However, often
specific model features, e.g. nonlinear equations of state  are included at the
level of the dynamic equations in an ad hoc fashion, neglecting
thereby the fact, that the passive surfactant-thin film system has to
respect symmetries imposed by the laws of thermodynamics
(see \cite{ThAP2012pf} for review).   

The recent formulation of the dynamic equations in terms of a
thermodynamically consistent gradient dynamics
\cite{ThAP2012pf,thap2016prf} sheds some light on a more
rigorous approach to model surfactant driven thin film flows using an
energy functional. Following the approach from Refs.
\cite{ThAP2012pf,thap2016prf}, features like nonlinear
equations of state for the surfactant and concentration-dependent
wettability can be included in a consistent manner into a mesoscopic
description. However, what still needs to be established is the
consistency of the mesoscopic approach with macroscopic parameters,
i.e. the equilibrium contact angle of a droplet in the presence of
surfactants. This relation has been derived by Sharma
\cite{Shar1993l} for droplets of pure liquids on a solid
substrate, by relating the mesoscopic parameters of the wetting energy
to the macroscopic Young-Dupr\'e equation and is e.g. also discussed in \cite{BMQ+1991l} for different wetting scenarios.

Here we establish this mesoscopic-macroscopic link for the extended
system: a droplet of a pure liquid in contact with a solid substrate
covered by a liquid adsorption layer in the presence of insoluble
surfactants. Our approach is based on a mesoscopic energy functional
depending on the film height and the surfactant
coverage profiles. 
We reveal the selection of the contact angle $\theta_e$ in the presence of surfactants. This  involves a nontrivial coupling with the equilibration of surfactant concentrations, respectively on the drop and on the liquid adsoprtion layer. 
These considerations are relevant for
cases involving bare substrates or ultra thin films, where apolar
and/or polar forces between interfaces become non-negligible and where
the dynamics is governed by the contact line. For example, it has been
proposed that the onset of Marangoni flows for surfactant driven
spreading and fingering of droplets on hydrophilic surfaces depends on
the ability of the surfactant to diffuse in front of the droplet to
establish a gradient, which then drives the flow
\cite{CCBV1999csa,CaCa1999l}. Similarly,
autophobing is associated with a transfer of surfactant onto the
substrate to render it less hydrophilic
\cite{AALM2004l,CrMa2007l}, leading to dewetting
and film rupture.  Although surfactant induced flows are dynamic
phenomena out of equilibrium, the underlying theoretical framework of
linear flux-force relations has to be consistent with the equilibrium
conditions at the meso- and macroscale.

The paper is structured as follows: First in section II, we will revise how to derive the macroscopic and mesoscopic equilibrium descriptions for a surfactant-free droplet of a pure liquid on a solid substrate. This parallel approach
establishes the link between the macroscopic variables (surfaces
tensions) and the additional mesoscopic variables (wetting energy) via
the Young-Dupr\'e law. While this section gives identical
results as Ref. \cite{Shar1993l}, it is nevertheless a
pedagogical introduction to the more involved calculations in the
presence of surfactants, which constitutes section III of the
paper. We rely here strictly on the existence of a (generalized)
Hamiltonian,
which includes capillarity and a wetting energy, both dependent on the
surfactant concentration.
No other assumptions about the underlying hydrodynamics of the problem
are made.  We show the conditions for consistency
between the macroscopic and mesoscopic approach in terms of the equilibrium
contact angle and the equilibrium distribution of surfactants. In
section IV, we illustrate our calculations by
explicitly choosing a functional form for the Hamiltonian, consistent
with a linear equation of state for the surfactant and we propose a
simple modification of the disjoining pressure which yields
consistency with the Young-Dupr\'e law in the presence of surfactants.  

\section{A drop of simple liquid (no surfactants)}
\subsection{Macroscopic consideration}
\label{sec:nma}
\begin{figure*}
\includegraphics[width = 0.8\textwidth]{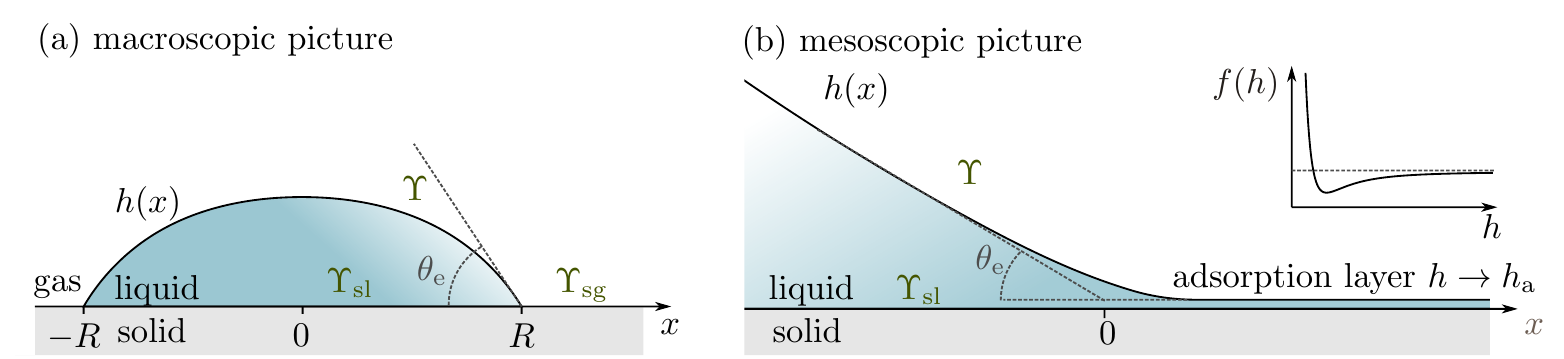}
\caption{Liquid drop at a solid-gas interface. (a) In the
    macroscopic picture, the equilibrium contact angle $\theta_\mathrm{e}$ is
    determined by the interfacial tensions $\Upsilon$, $\Upsilon_\mathrm{sl}$
    and $\Upsilon_\mathrm{sg}$, characterizing the liquid-gas, solid-liquid
    and solid-gas interface, respectively. (b) In the mesoscopic picture, the substrate is covered by an equilibrium adsorption layer of height $h_\mathrm{a}$ which corresponds to the minimum of the wetting energy $f(h)$.}
\label{Fig:Fig1}
\end{figure*}

We start by reviewing the derivation of the Laplace pressure and the
Young-Dupr\'e law from a free energy approach that we will later
expand by incorporating surfactants. Let us consider a 2D liquid drop
of finite volume, i.e., a cross section of a transversally invariant
liquid ridge, that has contact lines at $x=\pm R$ (see sketch
Fig.~\ref{Fig:Fig1} (a)). The liquid-gas, solid-liquid and solid-gas
free energy per area here directly correspond to the interface
tensions and are denoted by $\Upsilon$,
$\Upsilon_\mathrm{sl}$ and $\Upsilon_\mathrm{sg}$, respectively.
Using the drop's reflection symmetry the
(half) free energy is
\begin{equation}
\mathcal{F} = \int_0^R dx \, \left[ \Upsilon \xi+ \Upsilon_\mathrm{sl}
  - Ph \right]  + \int_R^\infty dx \, \Upsilon_\mathrm{sg} + \lambda_h
h(R). \label{no:s:macro}  
\end{equation}
where the metric factor is 
\begin{equation}
\xi=\left( 1+(\partial_x h)^2 \right)^{1/2} \label{metric:factor}
\end{equation}
and $\partial_x$ denotes the derivative w.r.t. $x$.
For
small interface slopes one can make the small-gradient or long-wave
approximation 
\begin{equation}
\xi\approx1+(\partial_x h)^2/2 \label{metric:factor:approx}
\end{equation}
often used in gradient
dynamics models on the interface Hamiltonian (aka thin-film or
lubrication models) \cite{OrDB1997rmp,Thie2010jpcm,YSTA2017pre}.
The liquid volume $V=\int dx h$ is controlled via the Lagrange
multiplier $P$.

We independently vary the profile $h(x)$ and the position of the
contact line $R$. The two are coupled due to $h(R)=0$, which is
imposed through the Lagrange multiplier $\lambda_h$. Varying $h(x)$ implies
\begin{equation}
\delta \mathcal{F} = \left[ \Upsilon\frac{\partial_x h}{\xi} + \lambda_h \right] \delta h(R)-
\int_0^R dx \,\delta h(x) \left[ \Upsilon \frac{\partial_{xx} h}{\xi^3} + P \right]
\end{equation}
which gives
\begin{eqnarray}
\lambda_h &=& -  \Upsilon\frac{\partial_{x} h}{\xi}, \quad {\rm for} \quad x=R, \label{eq:lamh}\\
P &=& -  \Upsilon \kappa, \quad {\rm for} \quad x \in [0,R],
\end{eqnarray}
where we introduced the curvature 
\begin{equation}
\kappa=\frac{\partial_{xx} h}{\xi^3} \, .
\end{equation}
The variation of $R$ evaluated at $x=R$ gives
\begin{equation}
\delta \mathcal{F} = \left[  \Upsilon \xi  + \Upsilon_\mathrm{sl} - \Upsilon_\mathrm{sg} - Ph  + \lambda_h (\partial_{x} h) \right] \delta R
\end{equation}
which together with the constraint $h(R)=0$ and $\lambda_h$
[Eq.~(\ref{eq:lamh})] results in the Young-Dupr\'e law
\begin{equation}
\Upsilon\cos\theta_\mathrm{e} = \Upsilon_\mathrm{sg} - \Upsilon_\mathrm{sl} \, ,
\label{eq:mayd}
\end{equation}
where we employed
\begin{equation}
1/\xi=\left( 1+(\partial_{x} h(R))^2 \right)^{-1/2}=\cos\theta_\mathrm{e}.  
\end{equation}
Note that a
similar approach is also presented in \cite{SA2008pof} and in \cite{Borm2009csaea}, where a
transversality condition at the boundary is used instead of a Lagrange
multiplier that fixes $h(R)=0$. 
Next, we remind the reader how to
obtain the same law from considerations on the mesoscale.
\subsection{Mesoscopic consideration}
\label{sec:nme}
Now we start from an interface Hamiltonian derived from microscopic 
considerations, asymptotically or numerically (see e.g., Refs. \cite{Diet1988,Schi1990,TMTT2013jcp,HuTA2017jcp})
\begin{equation}
\mathcal{F}= \int_{0}^\infty dx \, \left[ \Upsilon \xi +
  \Upsilon_\mathrm{sl} + f(h) - Ph \right] 
\label{eq-lagr}
\end{equation}
with the same metric factor defined in (\ref{metric:factor}). As in
(\ref{no:s:macro}) we consider only the half energy of a reflection symmetric droplet. Here $f(h)$ is the wetting potential \cite{Diet1988,Schi1990} as depicted in
Fig.~\ref{Fig:Fig1} (b). For partially wetting liquids $f(h)$
normally has a minimum at some $h=h_\mathrm{a}$ corresponding to the height of
an equilibrium adsorption layer (in hydrodynamics often referred to as
``precursor film'') and approaches zero as $h\to\infty$. 
Mathematically, $\mathcal{F}$ is a Lyapunov functional,
thermodynamically it may be seen as a grand potential, and in a
classical mechanical equivalent it would be an action (i.e., the
integral over the Lagrangian, with position $x$ and film height $h$ taking the roles of
time and position in classical point mechanics).

Now we vary $\mathcal{F}$ w.r.t.~$h(x)$ and obtain
\begin{equation}
\delta \mathcal{F} = \int_0^\infty dx \,\delta h(x) \left[-\Upsilon
  \kappa + \partial_h f - P \right]
\label{eq:nmedf}
\end{equation}
where we
used $[\Upsilon\frac{\partial_{x} h}{\xi}\delta h(x)]_0^\infty=0$.
Based on (\ref{eq:nmedf}), the free surface profile is given by the
Euler-Lagrange equation 
\begin{equation}
0 = -\Upsilon \kappa +\partial_h f -P\, .
\label{eq:nmep}
\end{equation}
Multiplying by $\partial_{x} h$ and integrating w.r.t.\ $x$ gives the 'first
integral'\footnote{%
Note, that if the integrand of (\ref{eq-lagr}) is seen as Lagrangian $L$, 
the generalized coordinate and corresponding momentum are $q=h$ and $p=\partial
L/\partial (\partial_x h) = \Upsilon (\partial_x h)/\xi$,
respectively. Then the first integral $E$
corresponds to the negative of the Hamiltonian $H=p\partial_x q - L$.
\label{foot-ham}}
\begin{align}
E &= -\Upsilon \int \frac{\partial_{x} h}{\xi^3} \partial_{xx} h \, dx + f(h) -P h  +\Upsilon_\mathrm{sl} \notag \\
&= \frac{\Upsilon}{\xi} + f(h) - P h +\Upsilon_\mathrm{sl} \,,
\label{eq:nmee}
\end{align}
where $E$ is a constant that is independent of $x$. This first integral can be interpreted as an energy density or as the horizontal force acting on a cross-section of the film. The fact that $E$ is constant reflects the horizontal force balance.

Now we consider the wedge geometry in
Fig.~\ref{Fig:Fig1}(b) and determine the thickness $h_\mathrm{a}$ of the coexisting
adsorption layer on the right and the angle $\theta_\mathrm{e}$ formed by the wedge on the
left. To do so, we first consider  Eqs.~(\ref{eq:nmep}) and (\ref{eq:nmee})
in the wedge region
far away from the adsorption layer, i.e., where the film height is sufficiently large that
$f, \partial_h f \to 0$ and $hP\to 0$. Note, that the mesoscopic wedge
region with $\partial_xh\approx \mathrm{const}$ is distinct from the region of the macroscopic droplet governed
by the Laplace law $P=-\Upsilon\kappa$ (For a more extensive argument
see Ref. \cite{Shar1993l}).
\\
This gives
\begin{eqnarray}
P &=& 0 \label{eq:nmepl}\\
E &=& \frac{\Upsilon}{\xi_\mathrm{w}}+\Upsilon_\mathrm{sl} \label{eq:nmeel}
\end{eqnarray}
in the wedge. Second, we consider the adsorption layer far away from the wedge. There, Eqs.~(\ref{eq:nmep}) and (\ref{eq:nmee}) result in
\begin{eqnarray}
P &=& \partial_h f|_{h_\mathrm{a}} \label{eq:nmepr}\\
E &=& \Upsilon + f(h_\mathrm{a}) - h_\mathrm{a} P + \Upsilon_\mathrm{sl} \, . \label{eq:nmeer}
\end{eqnarray}
Equilibrium states are characterized by a pressure $P$ and a first integral $E$ that are constant across the system. Therefore, the adsorption layer height $h_\mathrm{a}$ and the contact angle $\theta_\mathrm{e}$ are given by 
\begin{eqnarray}
 P  &=&  \partial_h f|_{h_\mathrm{a}} =0\quad \text{and}\\
 \frac{\Upsilon}{\xi_\mathrm{w}}&=&\Upsilon\cos\theta_\mathrm{e} = \Upsilon  + f(h_\mathrm{a}) \label{eq:nmeyd}
\end{eqnarray}
respectively.

\subsection{Consistency of mesoscopic and macroscopic approach}
Comparing Eq. (\ref{eq:nmeyd}) with the macroscopic Young-Dupr\'e law~(\ref{eq:mayd}) in section~\ref{sec:nma} yields the expected relation
\begin{equation}
f(h_\mathrm{a})=\Upsilon_\mathrm{sg}-\Upsilon_\mathrm{sl}-\Upsilon=S
\label{eq:cons}
\end{equation}
as condition for the consistency of mesoscopic and macroscopic
description. $S$ denotes the spreading coefficient. For small contact angles $\theta_\mathrm{e}\ll1$,
Eq.~(\ref{eq:nmeyd}) reads $f(h_\mathrm{a})=-\Upsilon\theta_\mathrm{e}^2/2$.

We can now reinterpret the free energy in (\ref{eq-lagr}).
The solid substrate with adsorption layer corresponds to the ``dry''
region in the macroscopic free energy (\ref{no:s:macro}). For
consistency at the energy level, the
mesoscopic energy density should approach
$\Upsilon_\mathrm{sg}$ in the adsorption layer at $P=0$ and
consequently $f(h_\mathrm{a}) = \Upsilon_\mathrm{sg}
-\Upsilon_\mathrm{sl} - \Upsilon$, which leads also to relation
(\ref{eq:cons}) \footnote{The solid substrate with adsorption layer corresponds to the ``moist case'' in
\cite{Genn1985rmp}, where the energy density should approach
$\Upsilon_\mathrm{sg}$ (strictly speaking
$\Upsilon_\mathrm{sg}^\mathrm{moist}$) and consequently $f(h_\mathrm{a}) = \Upsilon_\mathrm{sg}^\mathrm{moist} -
\Upsilon_\mathrm{sl} - \Upsilon$ as for a flat equilibrium adsorption
layer at $P=0$. This implies that the ``moist'' spreading coefficient is
$S^\mathrm{moist} = f(h_\mathrm{a})$ which is well defined as long as $f(h)$ has
a minimum. Note that for $h\to 0$, in many approximations the wetting energy
$f(h)$ shows an unphysical divergence. This may be avoided by
employing a cut-off (see e.g., \cite{Genn1985rmp,SA2008pof} or by determining
$f(h)$ from proper microscopic models
\cite{MacD2011epjst,TMTT2013jcp,HuTA2015jcp,HuTA2017jcp}).  In the
latter case one finds a finite
$f(0) = \Upsilon_\mathrm{sg}^\mathrm{dry} - \Upsilon_\mathrm{sl} -
\Upsilon=S^\mathrm{dry}$ well defined even for $f(h)$ without minimum.}.

As an aside, we note that
the here presented calculation is not exactly equivalent to the determination of a binodal for a binary
mixture where coexistence of two homogeneous phases is characterized by equal chemical
potential and equal local grand potential. Here, the coexistence of a
homogeneous phase (adsorption layer) and an inhomogeneous phase (wedge)
is characterized by equal pressure $P$ (corresponding to the chemical
potential in the case of a binary mixture) and equal Hamiltonian $E$ (which differs from
the local grand potential, i.e., the integrand in (\ref{eq-lagr}) by a factor $1/\xi^2$ in the liquid-gas
interface term).

\section{A liquid drop covered by insoluble surfactants}
\label{sec:s}
\subsection{Macroscopic consideration}
\label{sec:sma}
\begin{figure*}
\includegraphics[width = 0.8\textwidth]{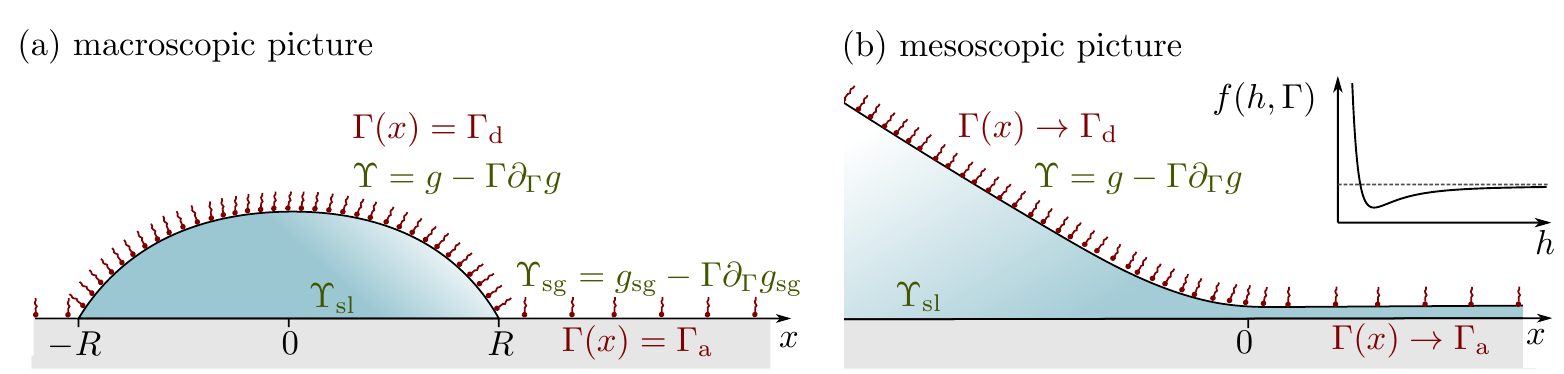}
\caption{Liquid drop covered by insoluble surfactant at a solid-gas interface. (a) In the macroscopic approach, the equilibrium contact angle is determined by the solid-liquid interfacial tension $\Upsilon_\mathrm{sl}$  and the interfacial tensions $\Upsilon$ and $\Upsilon_\mathrm{sg}$, which describe the liquid-gas and the the solid-gas interfacial tension  and which depend on the respective surfactant concentrations $\Gamma_\mathrm{d}$ and $\Gamma_\mathrm{a}$ on the droplet and the adsorption layer.
(b) In the mesoscopic picture, the substrate is covered by an equilibrium adsorption layer and the contact angle is determined by the liquid gas interfacial tension $\Upsilon$ which depends on the surfactant concentration,  the solid-liquid interfacial tension $\Upsilon_\mathrm{sl}$  and the minimum of the wetting energy $f(h_\mathrm{a})$. \label{Fig:Fig2}}
\end{figure*}

We now consider insoluble surfactants, which exhibit a number density
$\Gamma$ (per unit area) on the free liquid-gas interface $h(x)$ (see
Fig.~\ref{Fig:Fig2}(a)).  There may also be surfactant at the
solid-gas interface.  The total amount of surfactant, $N=\int ds \,
\Gamma= \int dx \xi \Gamma$, is conserved, which is imposed by a
Lagrange multiplier $\lambda_\Gamma$. The liquid volume $V=\int dx h$
and the condition $h(R)=0$ for a contact line at $R$ are ensured via
Lagrange multipliers $P$ and $\lambda_h$, respectively (as in
section~\ref{sec:nma}).  The surface free energies of the liquid-gas
and solid-gas interfaces are characterized by the functions
$g(\Gamma)$ and $g_\mathrm{sg}(\Gamma)$ respectively. The solid-liquid
interface is assumed to be free of surfactant.

As for the case of pure liquid in section~\ref{sec:nma}, we first consider a
macroscopic formulation in which the interaction of the
liquid-gas interface (and surfactants) with the
solid near the three-phase contact line is not made explicit -- this
is done in the mesoscopic model presented in
section~\ref{sec:surfmeso} below.

The energy now to be minimized corresponds to a grand potential and reads
\begin{align}
\mathcal{F}[h,\Gamma] =& \int_0^R dx \, \left[\xi g\left(\Gamma\right) + \Upsilon_\mathrm{sl} - Ph \right]  
+ \int_R^\infty dx \, g_\mathrm{sg}\left(\Gamma\right) \notag \\ 
& - \lambda_\Gamma \left(\int_0^R dx \, \xi \Gamma  + \int_R^\infty dx  \, \Gamma   \right)  + \lambda_h h(R).  
\label{eq:smacro_F}
\end{align}
Varying the field $\Gamma(x)$ gives
\begin{equation}
\delta{F} = \int_0^R dx \, \xi(\partial_\Gamma g -\lambda_{\Gamma} ) \delta \Gamma +\int_R^\infty dx \, (\partial_\Gamma g_\mathrm{sg} -\lambda_{\Gamma} ) \delta \Gamma
\end{equation}
resulting in
\begin{align}
& \lambda_\Gamma =\partial_\Gamma g \text{ for } x \in [0,R] \notag \\
\text{and} \quad  & \lambda_\Gamma =\partial_\Gamma g_\mathrm{sg} \text{ for } x \in [R,\infty]\,.
\label{eq:lgamm}
\end{align}
Since, in general, $\partial_\Gamma g$ is a function of $\Gamma$ and
$\lambda_\Gamma$ is a constant, Eq.~(\ref{eq:lgamm}) implies that
the surfactant is homogeneously distributed in each region, i.e.,
\begin{equation}
\partial_x\Gamma = 0.
\end{equation}
We introduce the equilibrium concentrations $\Gamma(x)=\Gamma_\mathrm{d}$ on the droplet and $\Gamma(x)=\Gamma_\mathrm{a}$ on the substrate.
For the equilibrium distribution of surfactants with constant chemical potential $\lambda_\Gamma$, Eq.~(\ref{eq:lgamm}) reduces to
\begin{equation}
\partial_\Gamma g|_{\Gamma_\mathrm{d}}=\partial_\Gamma g_\mathrm{sg}|_{\Gamma_\mathrm{a}}\,.\label{eq:lgamm:equ}
\end{equation}
Varying the field $h(x)$ gives
\begin{align}
\delta \mathcal{F} =  & 
\int_0^R dx \, \left[  -P - \frac{\partial_{xx} h }{\xi^3}(g-\lambda_\Gamma \Gamma  ) \right] \delta h(x) \notag \\
 & +   \left[ \left( \frac{\partial_x h}{\xi} (g - \lambda_\Gamma \Gamma) \right) \delta h \right]_0^R + \lambda_h \delta h(R) \notag \\
= &  \int_0^R dx \, \left[ -P -  \kappa \Upsilon  \right] \delta h(x) 
+\left[ \frac{\partial_x h}{\xi} \Upsilon + \lambda_h \right] \delta h(R)
\label{eq:varih}
\end{align}
where we employed Eq.~(\ref{eq:lgamm}) and introduced the 
surfactant-dependent liquid-gas interface tension (aka local grand potential,
aka mechanical tension in the interface or surface stress)
\begin{equation}
\Upsilon= g - \Gamma\partial_\Gamma g \, .
\label{eq:Ups}
\end{equation}
Note that indeed for insoluble surfactants a Wilhelmy plate in a Langmuir trough measures $\Upsilon$ and not $g$ as the area is changed at fixed amount of surfactant, i.e., $\Gamma$ changes with the area.
At the left boundary at $x=0$, the
reflection symmetry of the droplet enforces $\partial_x h=0$.
Eq.~(\ref{eq:varih}) implies that the Laplace pressure and $\lambda_h$ become
\begin{eqnarray}
P &=& -\Upsilon\kappa, \quad {\rm for} \quad x \in [0,R]\\
\lambda_h &=& -  \Upsilon\frac{\partial_x h}{\xi}, \quad {\rm at} \quad x=R.
\end{eqnarray}
Finally, variation of $R$ evaluated at $x=R$ gives:
\begin{align}
\delta \mathcal{F} =& [\xi   g\left(\Gamma\right) + \Upsilon_\mathrm{sl}- Ph  - g_\mathrm{sg}(\Gamma_\mathrm{a}) - \lambda_\Gamma 
  \xi\Gamma_\mathrm{d} \notag \\
  & - \lambda_\Gamma \Gamma_\mathrm{a} + \lambda_h \partial_x h(R)] \delta R
\end{align}
Using the constraint $h(R)=0$, as well as the obtained values for $\lambda_\Gamma$ and $\lambda_h$, this gives the boundary condition (using  $1/\xi=\cos\theta_\mathrm{e}$):
\begin{align}
0 & =\Upsilon_\mathrm{sl} - \Upsilon_\mathrm{sg}(\Gamma_\mathrm{a})
+ \Upsilon(\Gamma_\mathrm{d}) \cos\theta_\mathrm{e} \\
& \mathrm{with}\qquad 
\Upsilon(\Gamma_\mathrm{d})=g(\Gamma_\mathrm{d})-\Gamma_\mathrm{d} \partial_\Gamma g |_{\Gamma_\mathrm{d}}, \\
& \mathrm{and}\qquad \,
\Upsilon_\mathrm{sg}(\Gamma_\mathrm{a})=g_\mathrm{sg}(\Gamma_\mathrm{a})-\Gamma_\mathrm{a} \partial_\Gamma g_\mathrm{sg} |_{\Gamma_\mathrm{a}}\,, 
\label{eq:ydms}
\end{align}
i.e., we have again found the Young-Dupr\'e law that relates
interfacial tensions and equilibrium contact angle. However, the
interface tensions $\Upsilon_i$ are not based on the local free
energies $g$ and $g_\mathrm{sg}$ (which would enter at fixed concentration
$\Gamma$), but on the local grand potentials $g-\Gamma \partial_\Gamma
g$ and $g_\mathrm{sg}-\Gamma \partial_\Gamma g_\mathrm{sg}$ (valid at
constant total amount of surfactant).

Importantly, the values of $\Upsilon$ and $\Upsilon_{sg}$ are not fixed a priori, but have to be determined self-consistently from the equilibration of surfactant concentration, as given by (\ref{eq:lgamm:equ}). As such, the observed contact angle involves a subtle coupling between mechanics and distribution of surfactants. 

\subsection{Mesoscopic consideration}
\label{sec:surfmeso}
In analogy to section~\ref{sec:nme} where we developed mesoscopic considerations in the case without surfactant, next we discuss how to describe the case of insoluble surfactants on the mesoscale. Again we focus on equilibrium situations involving a contact line (Fig.~\ref{Fig:Fig2}(b)). Now it needs to be discussed how the dependency of the wetting potential on surfactant concentration has to be related to the respective dependencies of the involved surface energies to ensure consistency of mesoscopic and macroscopic descriptions.

A general discussion of a gradient dynamics description for the
dynamics of liquid layers or drops covered by insoluble or soluble
surfactants can be found in \cite{ThAP2012pf} and \cite{thap2016prf},
respectively. There, various thermodynamically consistent extensions
of thin film hydrodynamics without surfactants towards situations with
surfactants are discussed and contrasted to literature
approaches. Such extensions are, for instance, surfactant-dependent
interface energies and wetting potentials that affect not only
hydrodynamic flows but also diffusive
fluxes. However, the intrinsic relations between wetting energy $f$ and interface
energies $g$ were not discussed.

To begin with, we consider a general wetting
energy $f(h,\Gamma)$ and interface energies
$g(\Gamma)$. The resulting grand potential is
\begin{equation}
F[h, \Gamma] = \int_0^\infty \left[\Upsilon_\mathrm{sl}+f(h,\Gamma)+ g( \Gamma) \xi   - P h -\lambda_\Gamma
  \Gamma \xi \right] dx
\label{eq:surf-fen-j}
\end{equation}
with $\xi$ being again the metric factor (\ref{metric:factor}). $P$ and $\lambda_\Gamma$ are the Lagrange multipliers for the conservation of the amounts of liquid and surfactant, respectively. Note that we treat the solid-liquid interface energy $\Upsilon_\mathrm{sl}$ as constant. 

Varying $h(x)$ and $\Gamma(x)$ 
, we obtain from (\ref{eq:surf-fen-j}) the Euler-Lagrange equations
\begin{align}
 P&=  \partial_h f - \partial_x [ (g - \lambda_\Gamma \Gamma)
   \tfrac{\partial_x h}{\xi}  ]\label{eq:muh-j}\\
\mathrm{and}\qquad 
\lambda_\Gamma &= \frac{1}{\xi}\partial_\Gamma f + \partial_\Gamma g ,\label{eq:mugam-j}
\end{align}
respectively, i.e., the pressure $P$ and chemical potential
$\lambda_\Gamma$  are constant across the system. 

We use the mechanical approach of footnote~\ref{foot-ham} and introduce the generalized positions $q_1=h$ and $q_2=\Gamma$ and obtain from the local grand potential (integrand in (\ref{eq:surf-fen-j}), i.e., the 'Lagrangian'), the generalized momenta $p_1=(g-\lambda_\Gamma\Gamma)(\partial_x h)/\xi$ and $p_2=0$, respectively. In consequence, the first integral $E$ is
\begin{equation}
E=\Upsilon_\mathrm{sl} + f +\frac{g - \Gamma \lambda_\Gamma}{\xi} - h\,P 
\label{eq:surf-ham-j}
\end{equation}
i.e., Eq.~(\ref{eq:nmee}) with $\Upsilon$ replaced by $g(\Gamma) - \Gamma \lambda_\Gamma$.
All equilibrium states are characterized by $P$, $\lambda_\Gamma$ and
$E$ that are constant across the system. This allows us to investigate
the coexistence of states.

As in section \ref{sec:nme}, we consider the equilibrium between a 
wedge region with constant slope $\tan\theta_\mathrm{e}$
 and an adsorption layer of thickness $h_\mathrm{a}$ (Fig.~\ref{Fig:Fig2}(b)). As the wetting
potential $f(h,\Gamma)$ depends on film height and surfactant
concentration, one does not only need to determine the coexisting
wedge slope and adsorption layer height as in section~\ref{sec:nme} but also the coexisting surfactant
concentrations on the wedge, $\Gamma_\mathrm{w}$, and on the adsorption layer,
$\Gamma_\mathrm{a}$. 
The considered wedge is far away from the adsorption layer ($h\gg h_\mathrm{a}$, $f\to 0$, $|\partial_x h|\to \tan\theta_\mathrm{e}$, $\Gamma\to\Gamma_\mathrm{w}$) and the adsorption layer is far away from the wedge ($h\to h_\mathrm{a}, \partial_x h\to 0$, $\Gamma\to\Gamma_\mathrm{a}$), i.e., both are sufficiently far away from the contact line
region.  By comparing $P$, $\lambda_\Gamma$ and $E$  from Eqs.~(\ref{eq:muh-j}), (\ref{eq:mugam-j}) and
(\ref{eq:surf-ham-j}) in wedge and adsorption layer (in analogy to the calculation in section \ref{sec:nme}), one finds 
\begin{align}
0&=\partial_h f|_{(h_\mathrm{a}, \Gamma_\mathrm{a})},
\label{eq:mw1}\\
\partial_\Gamma g|_{\Gamma_\mathrm{w}} &= \partial_\Gamma f|_{(h_\mathrm{a}, \Gamma_\mathrm{a})} + \partial_\Gamma g|_{\Gamma_\mathrm{a}},
\label{eq:mw2}\\
\Upsilon(\Gamma_w) \cos\theta_\mathrm{e} &= f(h_\mathrm{a}, \Gamma_\mathrm{a}) - \Gamma_\mathrm{a} \partial_\Gamma f|_{(h_\mathrm{a}, \Gamma_\mathrm{a})} 
       +        \Upsilon(\Gamma_\mathrm{a}),
\label{eq:mw3}
\end{align}
respectively. To obtain
(\ref{eq:mw3}) we have already used (\ref{eq:mw1}) and (\ref{eq:mw2}) as well as
$\xi_\mathrm{w} = 1/\cos\theta_\mathrm{e}$ and (\ref{eq:Ups}). 
Without surfactant we recover Eq.~(\ref{eq:nmeyd}) of
section~\ref{sec:nme} as $g(0)$ is $\Upsilon$ of Sec. II.

The obtained Eqs.~(\ref{eq:mw1}) to (\ref{eq:mw3}) allow one to
determine the 'binodals' for the wedge-adsorption layer
coexistence. In practice, one may chose any of the four quantities
$\theta_\mathrm{e}, \Gamma_\mathrm{w}, h_\mathrm{a}, \Gamma_\mathrm{a}$ as
control parameter and determine the other three from the three
relations (\ref{eq:mw1})-(\ref{eq:mw3}). It is convenient to pick
$\Gamma_\mathrm{a}$ as control parameter and first use
Eq.~(\ref{eq:mw1}) to determine $h_\mathrm{a}$, then employ
Eq.~(\ref{eq:mw2}) to obtain $\Gamma_\mathrm{w}$ and, finally,
Eq.~(\ref{eq:mw3}) to get the equilibrium contact angle $\theta_\mathrm{e}$.
To obtain specific results, the wetting energy $f(h,\Gamma)$ and free
energies of the liquid-gas interface $g(\Gamma)$ and the solid-gas
  interface $g_\mathrm{sg}(\Gamma)$ have to be specified. A
simple but illustrative example is discussed in
section~\ref{sec:exampl}.

\subsection{Consistency of mesoscopic and macroscopic approach}
\label{sec:consmesomacro}
Eq.~(\ref{eq:mw3}) is the generalization of the mesoscopic Young-Dupr\'e law~(\ref{eq:nmeyd}) for the treated case with surfactant.
As the concentrations are different on the wedge ($\Gamma=\Gamma_\mathrm{w}$) and on the adsorption layer ($\Gamma=\Gamma_\mathrm{a}$), the liquid-gas interface tensions are also different.
Eq.~(\ref{eq:mw3}) is accompanied by Eqs.~(\ref{eq:mw1})
and~(\ref{eq:mw2}) that provide the adsorption
layer height and the relation between $\Gamma_\mathrm{w}$ and
$\Gamma_\mathrm{a}$, respectively.
Comparison of the mesoscopic Young-Dupr\'e law [Eq.~(\ref{eq:mw3})] with the macroscopic one [Eq.~(\ref{eq:ydms}) in section~\ref{sec:sma}] implies
\begin{equation}
f(h_\mathrm{a}, \Gamma_\mathrm{a})  -
\Gamma_\mathrm{a}\partial_\Gamma f|_{(h_\mathrm{a},
  \Gamma_\mathrm{a})} = \Upsilon_\mathrm{sg}(\Gamma_\mathrm{a})-\Upsilon_\mathrm{sl}
-\Upsilon(\Gamma_\mathrm{a})=S(\Gamma_\mathrm{a}).
    \label{eq:surf-yd3}
\end{equation}
This corresponds to a generalization of the consistency condition
(\ref{eq:cons}) for the case with surfactant. It relates the
macroscopic equations of state (or interface energies) with the
height- and surfactant-dependent wetting energy.\footnote{Note that alternatively one may instead of (\ref{eq:Ups}) define $\Upsilon = g - \Gamma \partial_\Gamma g - \Gamma/\xi \partial_\Gamma f $ rendering relations (\ref{eq:nmee}), (\ref{eq:nmeyd}), etc. formally valid at the cost of introducing a surfactant-, film height- and film slope-dependent surface tension.}

We have used that the surfactant concentrations
 should be identical in the macroscopic and the mesoscopic
 description. Note that the surfactant concentration $\Gamma_w$ on the wedge in the mesoscopic picture corresponds to the concentration $\Gamma_\mathrm{d}$ on the droplet in the macroscopic picture, as can be seen from Eq. (\ref{eq:mugam-j}). The consistency of the surfactant concentrations in both descriptions implies
another condition, namely, that the macroscopic chemical equilibrium
[Eq.~(\ref{eq:lgamm})] $\partial_\Gamma g|_{\Gamma_\mathrm{w}}
= \partial_\Gamma g_\mathrm{sg}|_{\Gamma_\mathrm{a}}$ has to coincide
with the mesoscopic one [Eq.~(\ref{eq:mw2})], i.e., $\partial_\Gamma
g|_{\Gamma_\mathrm{w}} = \partial_\Gamma f|_{(h_\mathrm{a},
  \Gamma_\mathrm{a})} + \partial_\Gamma
g|_{\Gamma_\mathrm{a}}$. Comparing the two conditions implies
\begin{equation}
\partial_\Gamma g_\mathrm{sg}|_{\Gamma_\mathrm{a}}=
\partial_\Gamma
f|_{(h_\mathrm{a}, \Gamma_\mathrm{a})} + \partial_\Gamma
g|_{\Gamma_\mathrm{a}}.
   \label{eq:mm-cons2}
\end{equation}
Introducing the resulting relation for $\partial_\Gamma
f|_{(h_\mathrm{a}, \Gamma_\mathrm{a})}$  into (\ref{eq:surf-yd3}) results in
\begin{equation}
f(h_\mathrm{a}, \Gamma_\mathrm{a})  = g_\mathrm{sg}(\Gamma_\mathrm{a})-\Upsilon_\mathrm{sl}
-g(\Gamma_\mathrm{a}).
    \label{eq:surf-yd3c}
\end{equation}

In the next section we explore the consequences of the consistency
conditions for a relatively simple case: First, we assume a low surfactant
concentration resulting in purely entropic interfacial
  energies $g(\Gamma)$ and $g_\mathrm{sg}(\Gamma)$ before extending the result to arbitrary $g$.

\section{Application for a simple energy}
\label{sec:exampl}

In the next section, we illustrate our examples for a simple free energy which describes the situation of a low concentration of surfactant.
We employ a wetting energy that is a product of height- and
concentration-dependent factors, i.e., the presence of surfactant only
changes the contact angle but not the adsorption layer height.
\subsection{Macroscopic consideration}
We consider a low concentration (ideal gas-like) insoluble surfactant on the solid-gas and the liquid-gas interface. In general, the surfactant will even in the dilute limit affect the liquid-gas and solid-gas interfaces differently, i.e., the relevant molecular
scales $a$ will differ due to different effective molecular areas. Thus we write the surface free energies $g(\Gamma)$ and $g_\mathrm{sg}(\Gamma)$ as
\begin{align}
 g(\Gamma) &= \Upsilon^0 + \frac{k_\mathrm{B}T}{a^2}\Gamma (\ln\Gamma -1) \label{eq:g_macro}\\
 g_\mathrm{sg}(\Gamma) &= \Upsilon_\mathrm{sg}^0+ \frac{k_\mathrm{B}T}{a_\mathrm{sg}^2}\Gamma (\ln\Gamma -1) \label{eq:gsg_macro}
\end{align}
respectively, i.e., introduce different effective molecular
length scales $a$ and $a_\mathrm{sg}$. This results in
\begin{align}
\Upsilon(\Gamma) &= g-\Gamma \partial_\Gamma g =\Upsilon^0 -\frac{k_\mathrm{B}T}{a^2}\Gamma \label{eq:Ups_macro} \\
\Upsilon_\mathrm{sg}(\Gamma) &= g_\mathrm{sg}-\Gamma \partial_\Gamma g_\mathrm{sg} =\Upsilon_\mathrm{sg}^0 -\frac{k_\mathrm{B}T}{a_\mathrm{sg}^2}\Gamma, \label{eq:Upssg_macro}
\end{align}
i.e., the purely entropic free energy results in a linear equation of state.
The macroscopic concentration-dependent $\Upsilon_\mathrm{sg}(\Gamma_\mathrm{a})$ reflects the fact
that the solid-gas interface is 'moist' as it is covered by the adsorption layer,
and at equilibrium, surfactant is found on the drop as well as on the
adsorption layer. As a result, the solid-gas interface tension $\Upsilon_\mathrm{sg}$ in
the macroscopic picture aggregates the effects of surfactant on
wetting energy \textit{and} interface energy $\Upsilon$.\\
By inserting these interface energies into the modified Young-Dupr\'e law (\ref{eq:ydms}), we find 
\begin{equation}
 \cos{\theta_\mathrm{e}}=\frac{\cos{\theta_{e0}}-\epsilon_1 \delta \Gamma_\mathrm{a}}{1-\epsilon_1\Gamma_\mathrm{d}}
 \label{eq_cos:_ex}
 \end{equation}
  with $\theta_{e0}$ being the contact angle in the absence of
  surfactant, $\delta= \frac{a^2}{a_\mathrm{sg}^2}$ being the ratio of the different molecular length scales and $\epsilon_1=k_\mathrm{B}T/(a^2\Upsilon^0)$ being a positive constant.
The ratio of surfactant concentrations follows directly from
Eqs. (\ref{eq:lgamm:equ}) as
\begin{equation}
\Gamma_\mathrm{d}=\Gamma_\mathrm{a}^\frac{a^2}{a_\mathrm{sg}^2} =\Gamma_\mathrm{a}^\delta  \, .\label{eq:gamratio_macro_ex}
\end{equation}
We discuss a number of limiting cases which distinguish between different ratios of the molecular length scales.
\begin{itemize}
 \item[(A)] The dependencies of the interface energies $g$ and
     $g_\mathrm{sg}$ on surfactant are identical, i.e., $a=a_\mathrm{sg}$ and therefore $\delta = {a^2}/{a_\mathrm{sg}^2}=1$.
      The surfactant concentrations on adsorption layer and
drop are identical ($\Gamma_\mathrm{d}=\Gamma_\mathrm{a}=\Gamma$).
The observable dependence of the equilibrium contact angle
  $\theta_\mathrm{e}$ on the surfactant concentration takes the form
  \begin{equation}
   \cos{\theta_\mathrm{e}}=\frac{\cos{\theta_{e0}}-\epsilon_1\Gamma}{1-\epsilon_1\Gamma}
  \end{equation}
    i.e. the contact angle would monotonically increase with the
    surfactant concentration, giving rise to the effect of autophobing.
\item[(B)]
The surfactant prefers to stay on the liquid-gas interface, i.e., $a\ll a_\mathrm{sg}$ and $\delta=a^2/a_\mathrm{sg}^2 \ll 1$. This implies
$\Gamma_\mathrm{d}\gg\Gamma_\mathrm{a}$. The equilibrium
contact angle shows the following functional dependence on the
surfactant concentration
\begin{equation}
\cos{\theta_\mathrm{e}} \approx \frac{\cos{\theta_{e0}}}{1-\epsilon_1\Gamma_\mathrm{d}} \,. 
\end{equation}
This case corresponds to the classical
surfactant effect, which decreases the equilibrium contact angle with
increasing concentration.
\item[(C)]
The surfactant prefers to stay on the solid-gas interface, i.e.,
$a\gg a_\mathrm{sg}$ and $\delta=a^2/a_\mathrm{sg}^2 \gg 1$, which implies
$\Gamma_\mathrm{d}\ll\Gamma_\mathrm{a}$. The equilibrium
contact angle shows the following functional dependence on the
surfactant concentration
\begin{equation}
\cos{\theta_\mathrm{e}}\approx \cos{\theta_{e0}}-\epsilon_1 \delta \Gamma_\mathrm{d} 
\end{equation}
This case corresponds to a strong autophobing effect, which increases the equilibrium contact angle with
increasing surfactant concentration.
\end{itemize}
These limiting cases illustrate that the dependency of $\theta_e$ with amount of surfactant depends subtly on the nature of the free energies. This will be further investigated numerically below in section \ref{subsec:numerics}.
\subsection{Mesoscopic consideration}
We again consider a low concentration (ideal gas-like) insoluble surfactant on the liquid-gas interface
with the ideal gas local free energy $g(\Gamma)$ as defined in (\ref{eq:g_macro}) and the liquid-gas interface tension $\Upsilon(\Gamma)$ as defined in (\ref{eq:Ups_macro}). Note that $g_\mathrm{sg}$ does not occur in the mesoscopic description as the whole domain is at least covered by an adsorption layer.
Further we use the strong assumption that the wetting energy
factorises as 
\begin{equation}
 f(h,\Gamma)=\chi(\Gamma) \hat f(h)
\label{eq:prodans}
\end{equation}
with $\chi(0)=1$.  This allows us to investigate the case of a
surfactant that influences the contact angle but does not change the
adsorption layer height.  The surfactant-independent adsorption layer
height $h_\mathrm{a}$ is still given by $\partial_h\hat f |_{h_\mathrm{a}}=P$ as
in section~\ref{sec:nme}. 
The equilibrium contact angle $\theta_\mathrm{e}$ is obtained by inserting the product
ansatz  (\ref{eq:prodans}) for $f(h, \Gamma)$ into (\ref{eq:mw3}), which results in
\begin{equation}
\Upsilon(\Gamma_\mathrm{w})\,\cos\theta_\mathrm{e} =
\Upsilon(\Gamma_\mathrm{a})\, + \hat f(h_\mathrm{a}) \left[\chi(\Gamma_\mathrm{a})  - \Gamma_\mathrm{a}
\partial_\Gamma \chi |_{\Gamma_\mathrm{a}}\right] \, .
    \label{eq:mw3_V2}
\end{equation}  

Note that the restriction to a simple
product ansatz implies that one is not able to investigate
surfactant-induced wetting transitions characterized by a diverging
adsorption layer height and we expect the ansatz to break down for
$\theta_\mathrm{e}\to0$.  This will be further discussed elsewhere.\footnote{%
  In general, it is known \cite{BEIM2009rmp} that two (independent)
  critical exponents characterize the change in wetting behavior
  close to the wetting transition: They characterize (i) how
  $\cos(\theta_\mathrm{e})$ approaches one and (ii) how the thickness of
  the adsorption layer diverges. Choosing a product ansatz
  corresponds to the limiting case of zero critical exponent for the
  adsorption layer height.  \\
}  

\subsection{Consistency of mesoscopic and macroscopic approach}
The concentration-dependence of $\chi(\Gamma)$ in (\ref{eq:prodans}) can not be chosen freely, but needs to account for the consistency condition of mesoscopic and macroscopic picture
(cf.~sections~\ref{sec:consmesomacro}). 
By inserting the product ansatz (\ref{eq:prodans}) for the wetting energy and the entropic local free energies into condition (\ref{eq:surf-yd3c}), which ensures the consistency of the two approaches, we obtain
\begin{align}
&\chi(\Gamma_\mathrm{a}) = 1 - M \Gamma_\mathrm{a} (\ln(\Gamma_\mathrm{a} -1) ) \notag \\
& \quad\mathrm{with}\quad M=\frac{k_\mathrm{B}T}{\hat
  f(h_\mathrm{a})}\left(\frac{1}{a^2}- \frac{1}{a_\mathrm{sg}^2}\right)\,.
\end{align}
As this expression has to hold for any $\Gamma_\mathrm{a}$, the wetting energy can be written as
\begin{equation}
f(h,\Gamma) =\hat f(h) \left[ 1 - \frac{k_\mathrm{B}T}{\hat
  f(h_\mathrm{a})}\left(\frac{1}{a^2}- \frac{1}{a_\mathrm{sg}^2}\right) \Gamma ( \ln \Gamma - 1)
\right] \, .
\label{eq:fdilute}
\end{equation}
Let us summarize the mesoscopic and the macroscopic approach for a drop covered by insoluble surfactant: Macroscopically, the situation is completely determined by $g$, $g_\mathrm{sg}$ and $\Upsilon_\mathrm{sl}$. This allows for given $\Gamma_\mathrm{a}$ or $\Gamma_\mathrm{d}$ to obtain the other $\Gamma$ and the contact angle $\theta_\mathrm{e}$.\\
Mesoscopically, $g_\mathrm{sg}$ is not defined, but via the consistency conditions it is reflected in the wetting energy $f(h,\Gamma)$ that itself is not part of the macroscopic description. In the special case
treated in this section, $g$ is determined by $a$, 
the macroscopic quantity $g_\mathrm{sg}$ is determined by $a_\mathrm{sg}$, und the concentration-dependence of the mesoscopic $f(h,\Gamma)$ depends on both, $a$ and $a_\mathrm{sg}$.
\begin{figure*}[tbh]
\includegraphics[width=0.4\textwidth]{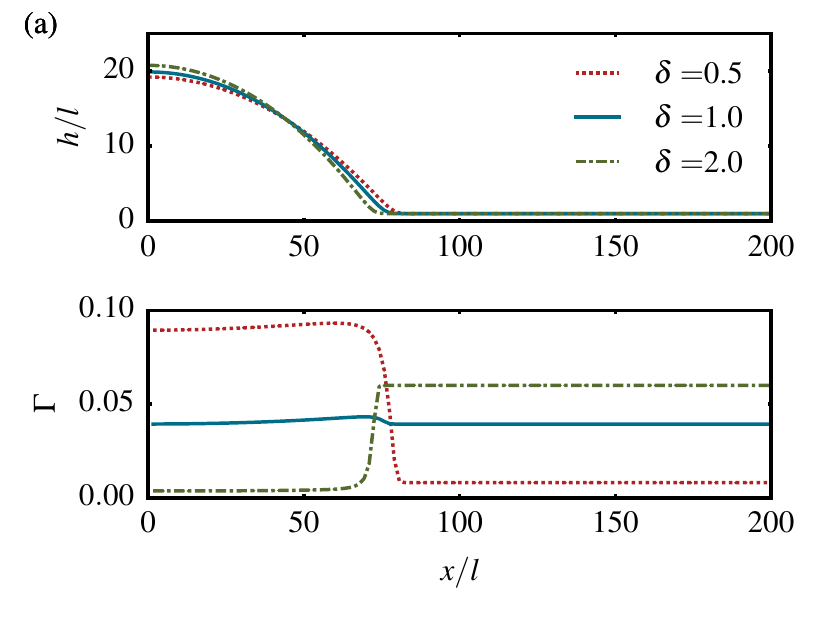}
\includegraphics[width=0.4\textwidth]{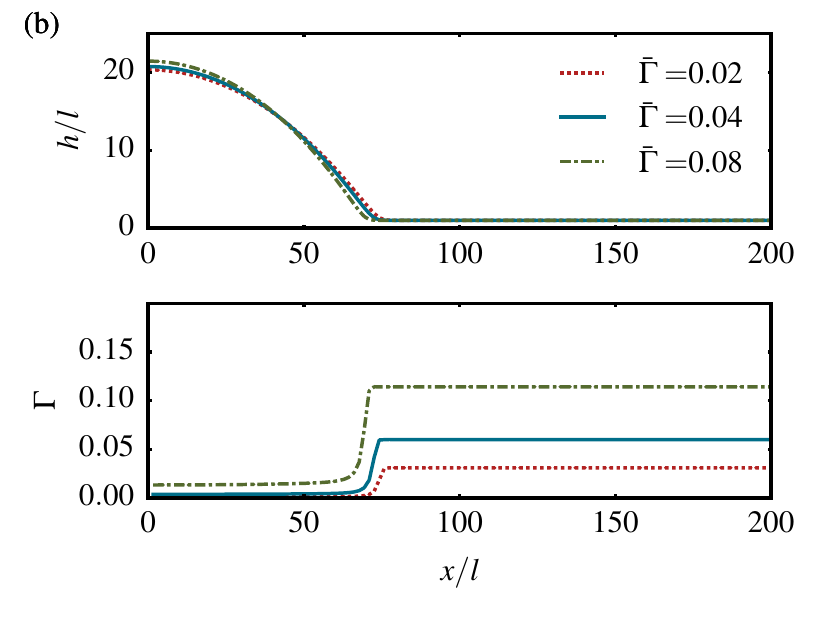}
\caption{Profiles of film height $h$ (top) and surfactant concentration $\Gamma$ (bottom) evolving in the numerical simulations for large times. The simulations are performed for three different ratios $\delta = \frac{a^2}{a_\mathrm{sg}^2}$ of the effective molecular length scales of the surfactant at $\bar{\Gamma}=0.04$ in (a) and three different mean surfactant concentrations $\bar{\Gamma}$ at $\delta=2$ in (b) while keeping the remaining parameters fixed to $\epsilon_1 = 0.2 $ and $\epsilon_2=0.4$. Note that the surfactant concentration $\Gamma_w$ which occurs on the wedge in the mesoscopic description corresponds to the concentration $\Gamma_\mathrm{d}$ on the droplet.}
\label{Fig3_Profiles}
\end{figure*}
\subsection{Numerical simulations for surfactant-covered drops on a finite domain}
\label{subsec:numerics}
To illustrate the equilibrium solutions of the model for finite domains,
we perform numerical time simulations of the evolution equations for film height and surfactant concentration. The emerging equilibrium states which arise in the time simulations at large times are then compared to the analytical predictions. 
As discussed in Refs.\cite{ThAP2012pf} and \cite{thap2016prf}, the evolution equations for a thin film covered by an insoluble surfactant can be written in the form of a gradient dynamics of the mesoscopic free energy functional $F$ given in Eq. (\ref{eq:surf-fen-j}) by introducing the projection of the surfactant concentration onto the flat surface of the substrate  $\tilde \Gamma = \xi \Gamma$
\begin{align}
\partial_t h &= \nabla \cdot [ Q_{hh} \nabla \frac{\delta F}{\delta h} + Q_{h\Gamma} \nabla \frac{\delta F}{\delta \tilde \Gamma}  ] \, , \\ 
\partial_t \tilde \Gamma &= \nabla \cdot [ Q_{\Gamma h} \nabla \frac{\delta F}{\delta h} + Q_{\Gamma \Gamma} \nabla \frac{\delta F}{\delta \tilde \Gamma}  ] \, ,
\label{eq:timeevolution} 
\end{align}
where the respective mobilities are denoted as $Q_{ij}$. In the following, we consider the wetting energy
 \begin{equation}
  f(h,\Gamma) =\chi(\Gamma) \hat f(h) 
 = \chi(\Gamma) \frac{A }{2 h ^2}\,\left(\frac{ 2h_\mathrm{a}^3 }{5 h^3} - 1 \right) \, ,
 \end{equation}
 where $\hat f(h)$ consists of two power laws and for $A>0$ describes a partially wetting fluid that macroscopically forms a droplet of finite contact angle on a stable adsorption layer of height $h_\mathrm{a}$.

 For the numerical analysis, the model is re-scaled, introducing the length scale $l=h_\mathrm{a}$. The solutions are characterized by three dimensionless parameters $\epsilon_1 = \frac{k_B T}{a^2 \Upsilon^0}$, $\epsilon_2 = - \frac{10 \hat{f}(h_\mathrm{a})}{3 \Upsilon_0}=\frac{A}{h_\mathrm{a}^2 \Upsilon_0}$ and $\delta= \frac{a^2}{a_\mathrm{sg}^2}$. These are connected to the ratio between the entropic contribution of the surfactant and the interfacial tension without surfactant, the equilibrium contact angle without surfactant and the ratio 
of the effective molecular length scales of the surfactant at the liquid-gas and solid-gas interface, respectively.
 
\begin{figure}[tbh]
\includegraphics[width=0.4\textwidth]{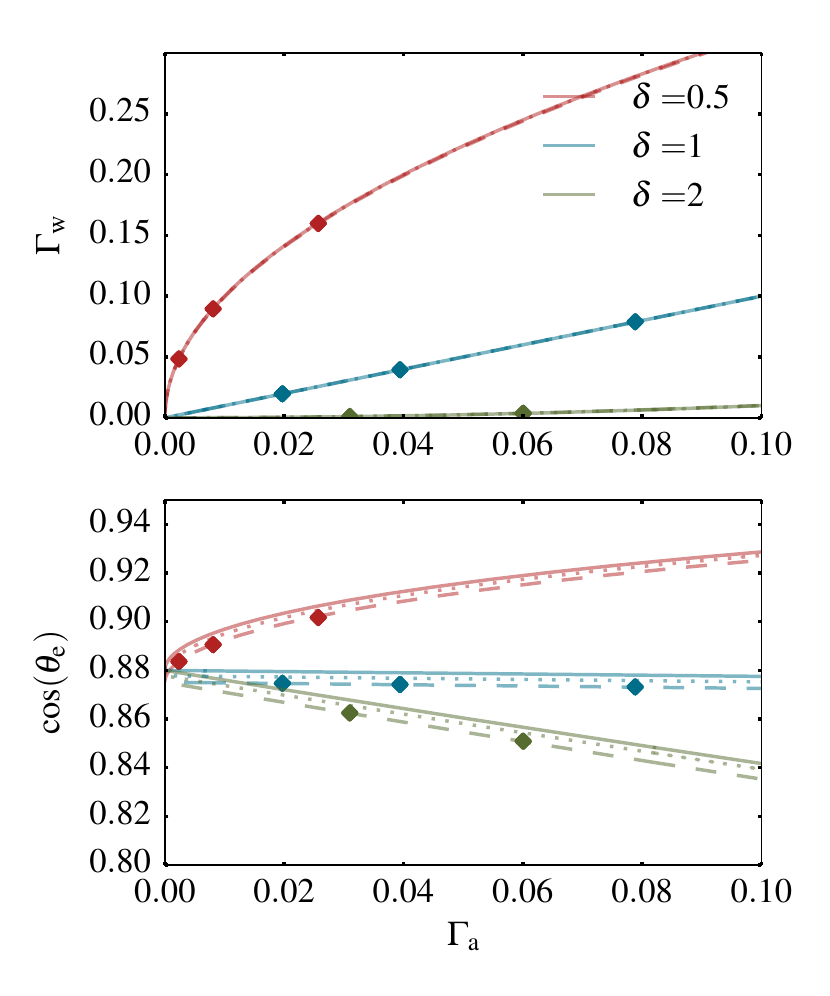}
\caption{Surfactant concentration on the droplet (top) and equilibrium contact angle (bottom) depending on the surfactant concentration in the adsorption layer. The analytically obtained equilibrium conditions (solid lines) are compared to values extracted from time simulations (diamonds) for three values of of $\delta$. The dashed (dotted lines) show the values obtained by parameter continuation for the domain size $L_x/l=200$ ($L_x/l=700$). The discrepancy in the equilibrium contact angle between the numerical and the analytical result can be attributed to finite size effects.}
\label{Fig4_Ratio_Angle}
\end{figure}

Starting with a droplet on an adsorption layer covered by a homogeneous surfactant concentration $\Gamma(x) = \bar{\Gamma}$ as initial condition, the 
evolution equations are solved using a finite element scheme provided by the modular toolbox DUNE-PDELAB \cite{BBD+2008C1, BBD+2008C2}. The simulation domain $\Omega = [0, L_x]$ with $L_x/l = 200$ is discretised on an equidistant mesh of $N_x=256$ quadratic elements with linear test and ansatz functions. No-flux boundary conditions are applied for both fields, corresponding to fixed amounts of fluid and surfactant in the system. For the time-integration, we employ an implicit Runge-Kutta scheme with adaptive time step and use the change in contact angle as the criterion to terminate the simulation when an equilibrium state is reached. 
\\
Figure \ref{Fig3_Profiles} shows the profiles for film height and surfactant concentration to which the system converges for large times. 
As examples, we study three different ratios $\delta$ of the effective molecular length scales of the surfactant while keeping the remaining parameters fixed to $\epsilon_1 = 0.2$ and $\epsilon_2=0.4$. The resulting profiles confirm the limiting cases discussed in section \ref{sec:exampl} A. If the dependencies of the interface energies $g$ and $g_\mathrm{sg}$ are identical, i.e. $a = a_\mathrm{sg}$ and thus $\delta = 1$ (solid blue lines), the surfactant concentration is identical on drop and adsorption layer. The addition of surfactant to the system has in this case only little effect on the contact angle. If the surfactant prefers to stay on the liquid-gas interface [$a<a_\mathrm{sg}$ and thus $\delta <1$ (dashed red lines)], the surfactant accumulates on the droplet and the contact angle is slightly lowered.  If the surfactant prefers to stay on the solid-gas interface [$a>a_\mathrm{sg}$ and thus $\delta >1$ (dash-dotted green lines)], the 
surfactant concentration on the drop is smaller then on the adsorption layer and the contact angle of the droplet increases. 
From the numerical time simulations, we extract the surfactant concentrations on the adsorption layer and on the droplet as well as the equilibrium contact angle and compare it to the analytically obtained equilibrium conditions (\ref{eq_cos:_ex}) and (\ref{eq:gamratio_macro_ex}) using the surfactant concentration on the adsorption layer $\Gamma_\mathrm{a}$ as control parameter. In the time simulations, different amounts of surfactant are simply implemented by changing the initial concentration of surfactant $\bar{\Gamma}$. 
Figure \ref{Fig4_Ratio_Angle} shows for three different values of $\delta$ the analytically obtained equilibrium values (\ref{eq_cos:_ex}) and (\ref{eq:gamratio_macro_ex}) depending on $\Gamma_\mathrm{a}$ as solid lines and the values extracted from time simulations with $L_x/l=200$ (diamonds). The surfactant concentrations measured in the time simulation (top) match the analytical prediction very well. However, there is a small discrepancy for the contact angles (bottom).
In order to understand this offset and test the hypothesis that it can be attributed to finite size effects, we analyse the steady state solutions using parameter continuation \cite{DWC+2014ccp} employing the software package AUTO-07p \cite{DOC+2007}.
The dashed lines in Figure \ref{Fig4_Ratio_Angle} show the concentration $\Gamma_w$ and $\cos(\theta_\mathrm{e})$ obtained by parameter continuation for a domain and droplet size that correspond to the values used in the time simulations. 
If the domain size is increased to $L_x/l=700$ with accordingly adjusted liquid volume, the values obtained by parameter continuation (dotted lines) are very close to the analytical prediction. The observed deviation of the time simulations can thus be explained by the finite size of the simulation domain and droplet. For very large domain and droplet sizes, the analytical predictions for surfactant concentration and contact angle perfectly match.
\subsection{Generalization to arbitrary interface energies}
Having established the form of the function $\chi(\Gamma)$ which guarantees
the consistency of the macroscopic and mesoscopic approach for the
equilibrium contact angle we can now write a free energy on the
mesoscale which is consistent with the macroscale. Identifying $\chi$
with
\begin{equation}
\chi=\tfrac{1}{\hat f_\mathrm{a}}\left[g_\mathrm{sg}(\Gamma)-g(\Gamma)\right]
\label{eq:generalchi}
\end{equation}
we can rewrite Eq.\ (\ref{eq:surf-fen-j}) as
\begin{align}
F[h,\Gamma]=& \int \Big\{ \Upsilon_\mathrm{sl}+\frac{\hat f(h)}{\hat f_\mathrm{a}}\left[g_\mathrm{sg}(\Gamma)-g(\Gamma)\right] \notag \\ 
&  + \xi\left[g(\Gamma)-\lambda\Gamma\right]-Ph \Big\} dx\,.\label{eq:surf-fen-j-meso-macro}
\end{align}
We split now the energy functional into three contributions stemming
from the droplet $F_\mathrm{drop}$, the contact line region $F_\mathrm{int}$ and the
adsorption layer $F_\mathrm{a}$, i.e. $F=F_\mathrm{drop}+F_\mathrm{int}+F_\mathrm{a}$. In the droplet, away from the contact line
(\ref{eq:surf-fen-j-meso-macro}) simplifies to 
\begin{equation}
F_{\mathrm{drop}}=\int\left\{\Upsilon_\mathrm{sl}+\xi\left[g(\Gamma)-\lambda\Gamma\right]-Ph\right\} dx\,,\label{eq:surf-f_en-j-w-meso}
\end{equation}
whereas in the adsorption layer we find
\begin{equation}
F_{\mathrm{a}}=\int\left\{\Upsilon_\mathrm{sl}+g_\mathrm{sg}-\lambda\Gamma\right\} dx\,. \label{eq:surf-f_en-j-a-meso}
\end{equation}
where we have dropped the pressure term $Ph_\mathrm{a}$ in $F_{\mathrm{a}}$ by
assuming that outside the adsorption layer $h\gg h_\mathrm{a}$ and that the
volume constraint on the liquid is determined by the droplet and not
the adsorption layer. Expressions (\ref{eq:surf-f_en-j-w-meso}) and
(\ref{eq:surf-f_en-j-a-meso}) are now identical to the macroscopic
description in section \ref{sec:sma} Eq. (\ref{eq:smacro_F})
This shows that the expression for $\chi(\Gamma)$ given in (\ref{eq:generalchi}) is valid for all expressions $g$ if the product ansatz for $f(h,\Gamma)$ is used.

\section{Conclusion and outlook}
We have employed equilibrium considerations to establish the link
between mesoscopic and macroscopic descriptions of drops covered by
insoluble surfactants that rest on smooth solid substrates. The
requirement of consistency of the two approaches relates the
macroscopic quantities (interface tensions) and the mesoscopic
quantities (wetting energy) and implies that the dependencies of
interface and wetting energies on surfactant concentration may not be
chosen independently. In particular the solid-gas interface tension in
the macroscopic description is directly related to properties of the
mesoscopic wetting energy.

The main conclusions of our equilibrium results also apply to the
theoretical description of out-of-equilibrium phenomena through
hydrodynamic modelling. In particular, the surfactant-dependencies of
Derjaguin (or disjoining) pressures and interface tensions may not be
chosen independently as this might result in (i) incorrect dynamics
towards equilibrium and (ii) incorrect final states, i.e., that do not
correspond to minima of appropriate energy functionals. We emphasize
that although many phenomena associated with surfactants, like
autophobing or spreading, are typically studied in dynamic and out of
equilibrium settings, an underlying mesoscopic theoretical framework
should for large times  always lead to the same equilibrium state as the corresponding macroscopic description.

If one does not take the consistency relation into account and chooses in the mesoscopic
model the surfactant-dependencies of liquid-gas interface tension and
wetting energy without having the macroscopic system in mind one may
implicitly assume quite peculiar surfactant-dependencies of
the solid-gas interface tension.\footnote{For instance, the linear dependencies
  of the Hamaker and liquid-gas interface tension on surfactant
  concentration employed in section V.C.1 of \cite{WaCM2002pof} imply
a solid gas interface tension of the form
$c_1+c_2\Gamma+c_3(1+\Gamma)^{n/(m-n)})$ where the $c_i$ are constants and $n$ and $m$ are the powers in a polynomial wetting energy.}

In section~\ref{sec:exampl}, we have used a specific simple
example to illustrate how the wetting energy (and Derjaguin pressure) needs to be
modified in the presence of surfactants with a linear equation of
state to ensure consistency between the macroscopic and the mesoscopic
picture. Note that the employed ansatz of a factorized wetting energy
$f(h)= \hat{f}(h) \chi(\Gamma)$ was chosen for simplicity. It is just
one possible choice and actually strongly restricts the physical
phenomena that can be described. To model, e.g. the behaviour close to
a wetting transition, other assumptions regarding the form of the
wetting energy need to be made as the product ansatz fixes the height
of the adsorption layer while at a wetting transition it diverges.

The main arguments and results of our work as detailed in
section~\ref{sec:s} are, however, of a general nature. They are
independent of the exact form of the wetting energy.  We find that in
the presence of surfactant, the structural form of the Young-Dupr{\'e}
law remains unchanged, but the surfactant concentrations and surface
tensions equilibrate self-consistently. Depending on the relation of
the interface free energies of liquid-gas and solid-gas interfaces,
adding surfactant may have qualitatively different effects on the
contact angle. Even in our simple example with purely entropic
surfactant free energies, we either find a lowering of the contact
angle with increasing amount of surfactant in the system or the
opposite behaviour, i.e., an autophobing. The approach proposed here
together with the general dynamic models introduced in
\cite{ThAP2012pf,thap2016prf} allows for systematic numerical
investigations of drop spreading and retraction dynamics employing
mesoscopic models with consistent dependencies of wetting energy and
interface tensions on surfactant concentration. For overviews of rich
spreading, autophobing and fingering behaviour in various experiments
see e.g.\
\cite{MaLe1981cec,FrGa1995l,BCCF1996fd,Stoe1996l,StKV2000jcis,SKSC2012l}.

As our approach is generic it may be extended to a number of more
complex situations. For instance, the surfactant can accumulate at all
three interfaces. Then in the macroscopic picture, the liquid-gas,
solid-liquid and solid-gas interfaces are characterized by
surfactant-dependent local free energies $g(\Gamma)$,
$g_\mathrm{sl}(\Gamma)$, and $g_\mathrm{sg}(\Gamma)$,
respectively. For a fixed overall amount of surfactant one again
obtains Eq.~(\ref{eq:ydms}), however, all interface tensions
correspond to local grand potentials:
$\Upsilon_i=g_i(\Gamma_i)-\Gamma_i\partial_{\Gamma}g_i |_{\Gamma_i}$
and the three concentrations $\Gamma_i$ are related by
$\partial_{\Gamma} g|_{\Gamma=\Gamma_\mathrm{lg}}=\partial_{\Gamma}
g_\mathrm{sl}|_{\Gamma=\Gamma_\mathrm{sl}}=\partial_{\Gamma}
g_\mathrm{sg}|_{\Gamma=\Gamma_\mathrm{sg}}$, i.e., the chemical
potential is uniform across the entire system. The incorporation of
surfactant also at the solid-liquid interface thus renders the
discussion more involved but does not pose a principal problem as long
as such a dependency is also incorporated into the mesoscale
consideration. A further example is a generalization towards soluble
surfactants. Then, additionally a bulk concentrations of surfactants
has to be incorporated in the static consideration (for fully dynamic
thin-film models see \cite{thap2016prf}). Incorporation of micelles is
also possible.

In principle, the local free energies (or equation of state) for the
surfactant may be arbitrarily complicated and account e.g., for phase
transitions of the surfactant. This can include substrate-induced
phase transitions as substrate-mediated condensation
\cite{RiSp1992tsf,KGFT2012njp}. If such transitions can occur the free
energy also has to account for gradient contributions in the
surfactant concentration (see e.g., extensions discussed in
\cite{ThAP2012pf,thap2016prf}). The approach developed here would then
again give consistency relations between interface and wetting energies
and include the possibility of phase changes in the surfactant layer.


\begin{thebibliography}{42}%
\makeatletter
\providecommand \@ifxundefined [1]{%
 \@ifx{#1\undefined}
}%
\providecommand \@ifnum [1]{%
 \ifnum #1\expandafter \@firstoftwo
 \else \expandafter \@secondoftwo
 \fi
}%
\providecommand \@ifx [1]{%
 \ifx #1\expandafter \@firstoftwo
 \else \expandafter \@secondoftwo
 \fi
}%
\providecommand \natexlab [1]{#1}%
\providecommand \enquote  [1]{``#1''}%
\providecommand \bibnamefont  [1]{#1}%
\providecommand \bibfnamefont [1]{#1}%
\providecommand \citenamefont [1]{#1}%
\providecommand \href@noop [0]{\@secondoftwo}%
\providecommand \href [0]{\begingroup \@sanitize@url \@href}%
\providecommand \@href[1]{\@@startlink{#1}\@@href}%
\providecommand \@@href[1]{\endgroup#1\@@endlink}%
\providecommand \@sanitize@url [0]{\catcode `\\12\catcode `\$12\catcode
  `\&12\catcode `\#12\catcode `\^12\catcode `\_12\catcode `\%12\relax}%
\providecommand \@@startlink[1]{}%
\providecommand \@@endlink[0]{}%
\providecommand \url  [0]{\begingroup\@sanitize@url \@url }%
\providecommand \@url [1]{\endgroup\@href {#1}{\urlprefix }}%
\providecommand \urlprefix  [0]{URL }%
\providecommand \Eprint [0]{\href }%
\providecommand \doibase [0]{http://dx.doi.org/}%
\providecommand \selectlanguage [0]{\@gobble}%
\providecommand \bibinfo  [0]{\@secondoftwo}%
\providecommand \bibfield  [0]{\@secondoftwo}%
\providecommand \translation [1]{[#1]}%
\providecommand \BibitemOpen [0]{}%
\providecommand \bibitemStop [0]{}%
\providecommand \bibitemNoStop [0]{.\EOS\space}%
\providecommand \EOS [0]{\spacefactor3000\relax}%
\providecommand \BibitemShut  [1]{\csname bibitem#1\endcsname}%
\let\auto@bib@innerbib\@empty
\bibitem [{\citenamefont {Craster}\ and\ \citenamefont
  {Matar}(2009)}]{CrMa2009rmp}%
  \BibitemOpen
  \bibfield  {author} {\bibinfo {author} {\bibfnamefont {R.}~\bibnamefont
  {Craster}}\ and\ \bibinfo {author} {\bibfnamefont {O.}~\bibnamefont
  {Matar}},\ }\href@noop {} {\bibfield  {journal} {\bibinfo  {journal} {Rev.
  Mod. Phys.}\ }\textbf {\bibinfo {volume} {81}},\ \bibinfo {pages} {1131}
  (\bibinfo {year} {2009})}\BibitemShut {NoStop}%
\bibitem [{\citenamefont {Matar}\ and\ \citenamefont
  {Craster}(2009)}]{MaCr2009sm}%
  \BibitemOpen
  \bibfield  {author} {\bibinfo {author} {\bibfnamefont {O.}~\bibnamefont
  {Matar}}\ and\ \bibinfo {author} {\bibfnamefont {R.}~\bibnamefont
  {Craster}},\ }\href@noop {} {\bibfield  {journal} {\bibinfo  {journal} {Soft
  Matter}\ }\textbf {\bibinfo {volume} {5}},\ \bibinfo {pages} {3801} (\bibinfo
  {year} {2009})}\BibitemShut {NoStop}%
\bibitem [{\citenamefont {Marmur}\ and\ \citenamefont
  {Lelah}(1981)}]{MaLe1981cec}%
  \BibitemOpen
  \bibfield  {author} {\bibinfo {author} {\bibfnamefont {A.}~\bibnamefont
  {Marmur}}\ and\ \bibinfo {author} {\bibfnamefont {M.~D.}\ \bibnamefont
  {Lelah}},\ }\href {\doibase 10.1080/00986448108910901} {\bibfield  {journal}
  {\bibinfo  {journal} {Chem. Eng. Commun.}\ }\textbf {\bibinfo {volume}
  {13}},\ \bibinfo {pages} {133} (\bibinfo {year} {1981})}\BibitemShut
  {NoStop}%
\bibitem [{\citenamefont {Troian}\ \emph {et~al.}(1989)\citenamefont {Troian},
  \citenamefont {Wu},\ and\ \citenamefont {Safran}}]{TrWS1989prl}%
  \BibitemOpen
  \bibfield  {author} {\bibinfo {author} {\bibfnamefont {S.}~\bibnamefont
  {Troian}}, \bibinfo {author} {\bibfnamefont {X.}~\bibnamefont {Wu}}, \ and\
  \bibinfo {author} {\bibfnamefont {S.}~\bibnamefont {Safran}},\ }\href@noop {}
  {\bibfield  {journal} {\bibinfo  {journal} {Phys. Rev. Lett.}\ }\textbf
  {\bibinfo {volume} {62}},\ \bibinfo {pages} {1496} (\bibinfo {year}
  {1989})}\BibitemShut {NoStop}%
\bibitem [{\citenamefont {Troian}\ \emph {et~al.}(1990)\citenamefont {Troian},
  \citenamefont {Herbolzheimer},\ and\ \citenamefont {Safran}}]{TrHS1990prl}%
  \BibitemOpen
  \bibfield  {author} {\bibinfo {author} {\bibfnamefont {S.}~\bibnamefont
  {Troian}}, \bibinfo {author} {\bibfnamefont {E.}~\bibnamefont
  {Herbolzheimer}}, \ and\ \bibinfo {author} {\bibfnamefont {S.}~\bibnamefont
  {Safran}},\ }\href@noop {} {\bibfield  {journal} {\bibinfo  {journal} {Phys.
  Rev. Lett.}\ }\textbf {\bibinfo {volume} {65}},\ \bibinfo {pages} {333}
  (\bibinfo {year} {1990})}\BibitemShut {NoStop}%
\bibitem [{\citenamefont {Cachile}\ \emph {et~al.}(1999)\citenamefont
  {Cachile}, \citenamefont {Cazabat}, \citenamefont {Bardon}, \citenamefont
  {Valignat},\ and\ \citenamefont {Vandenbrouck}}]{CCBV1999csa}%
  \BibitemOpen
  \bibfield  {author} {\bibinfo {author} {\bibfnamefont {M.}~\bibnamefont
  {Cachile}}, \bibinfo {author} {\bibfnamefont {A.}~\bibnamefont {Cazabat}},
  \bibinfo {author} {\bibfnamefont {S.}~\bibnamefont {Bardon}}, \bibinfo
  {author} {\bibfnamefont {M.}~\bibnamefont {Valignat}}, \ and\ \bibinfo
  {author} {\bibfnamefont {F.}~\bibnamefont {Vandenbrouck}},\ }\href@noop {}
  {\bibfield  {journal} {\bibinfo  {journal} {Colloids Surf., A}\ }\textbf
  {\bibinfo {volume} {159}},\ \bibinfo {pages} {47} (\bibinfo {year}
  {1999})}\BibitemShut {NoStop}%
\bibitem [{\citenamefont {Cachile}\ and\ \citenamefont
  {Cazabat}(1999)}]{CaCa1999l}%
  \BibitemOpen
  \bibfield  {author} {\bibinfo {author} {\bibfnamefont {M.}~\bibnamefont
  {Cachile}}\ and\ \bibinfo {author} {\bibfnamefont {A.}~\bibnamefont
  {Cazabat}},\ }\href@noop {} {\bibfield  {journal} {\bibinfo  {journal}
  {Langmuir}\ }\textbf {\bibinfo {volume} {15}},\ \bibinfo {pages} {1515}
  (\bibinfo {year} {1999})}\BibitemShut {NoStop}%
\bibitem [{\citenamefont {Hill}(1998)}]{Hill1998coc}%
  \BibitemOpen
  \bibfield  {author} {\bibinfo {author} {\bibfnamefont {R.}~\bibnamefont
  {Hill}},\ }\href@noop {} {\bibfield  {journal} {\bibinfo  {journal} {Current
  opinion in colloid \& interface science}\ }\textbf {\bibinfo {volume} {3}},\
  \bibinfo {pages} {247} (\bibinfo {year} {1998})}\BibitemShut {NoStop}%
\bibitem [{\citenamefont {Rafa{\"\i}}\ \emph {et~al.}(2002)\citenamefont
  {Rafa{\"\i}}, \citenamefont {Sarker}, \citenamefont {Bergeron}, \citenamefont
  {Meunier},\ and\ \citenamefont {Bonn}}]{RSBM2002l}%
  \BibitemOpen
  \bibfield  {author} {\bibinfo {author} {\bibfnamefont {S.}~\bibnamefont
  {Rafa{\"\i}}}, \bibinfo {author} {\bibfnamefont {D.}~\bibnamefont {Sarker}},
  \bibinfo {author} {\bibfnamefont {V.}~\bibnamefont {Bergeron}}, \bibinfo
  {author} {\bibfnamefont {J.}~\bibnamefont {Meunier}}, \ and\ \bibinfo
  {author} {\bibfnamefont {D.}~\bibnamefont {Bonn}},\ }\href@noop {} {\bibfield
   {journal} {\bibinfo  {journal} {Langmuir}\ }\textbf {\bibinfo {volume}
  {18}},\ \bibinfo {pages} {10486} (\bibinfo {year} {2002})}\BibitemShut
  {NoStop}%
\bibitem [{\citenamefont {Afsar-Siddiqui}\ \emph {et~al.}(2004)\citenamefont
  {Afsar-Siddiqui}, \citenamefont {Luckham},\ and\ \citenamefont
  {Matar}}]{AALM2004l}%
  \BibitemOpen
  \bibfield  {author} {\bibinfo {author} {\bibfnamefont {A.}~\bibnamefont
  {Afsar-Siddiqui}}, \bibinfo {author} {\bibfnamefont {P.}~\bibnamefont
  {Luckham}}, \ and\ \bibinfo {author} {\bibfnamefont {O.}~\bibnamefont
  {Matar}},\ }\href {\doibase 10.1021/la040041z} {\bibfield  {journal}
  {\bibinfo  {journal} {Langmuir}\ }\textbf {\bibinfo {volume} {20}},\ \bibinfo
  {pages} {7575} (\bibinfo {year} {2004})}\BibitemShut {NoStop}%
\bibitem [{\citenamefont {Craster}\ and\ \citenamefont
  {Matar}(2007)}]{CrMa2007l}%
  \BibitemOpen
  \bibfield  {author} {\bibinfo {author} {\bibfnamefont {R.}~\bibnamefont
  {Craster}}\ and\ \bibinfo {author} {\bibfnamefont {O.}~\bibnamefont
  {Matar}},\ }\href {\doibase 10.1021/la0629936} {\bibfield  {journal}
  {\bibinfo  {journal} {Langmuir}\ }\textbf {\bibinfo {volume} {23}},\ \bibinfo
  {pages} {2588} (\bibinfo {year} {2007})}\BibitemShut {NoStop}%
\bibitem [{\citenamefont {Bera}\ \emph {et~al.}(2016)\citenamefont {Bera},
  \citenamefont {Duits}, \citenamefont {Stuart}, \citenamefont {van~den Ende},\
  and\ \citenamefont {Mugele}}]{BDSE2016sm}%
  \BibitemOpen
  \bibfield  {author} {\bibinfo {author} {\bibfnamefont {B.}~\bibnamefont
  {Bera}}, \bibinfo {author} {\bibfnamefont {M.}~\bibnamefont {Duits}},
  \bibinfo {author} {\bibfnamefont {M.}~\bibnamefont {Stuart}}, \bibinfo
  {author} {\bibfnamefont {D.}~\bibnamefont {van~den Ende}}, \ and\ \bibinfo
  {author} {\bibfnamefont {F.}~\bibnamefont {Mugele}},\ }\href {\doibase
  10.1039/c6sm00128a} {\bibfield  {journal} {\bibinfo  {journal} {Soft Matter}\
  }\textbf {\bibinfo {volume} {12}},\ \bibinfo {pages} {4562} (\bibinfo {year}
  {2016})}\BibitemShut {NoStop}%
\bibitem [{\citenamefont {Thiele}\ \emph {et~al.}(2012)\citenamefont {Thiele},
  \citenamefont {Archer},\ and\ \citenamefont {Plapp}}]{ThAP2012pf}%
  \BibitemOpen
  \bibfield  {author} {\bibinfo {author} {\bibfnamefont {U.}~\bibnamefont
  {Thiele}}, \bibinfo {author} {\bibfnamefont {A.~J.}\ \bibnamefont {Archer}},
  \ and\ \bibinfo {author} {\bibfnamefont {M.}~\bibnamefont {Plapp}},\ }\href
  {\doibase 10.1063/1.4758476} {\bibfield  {journal} {\bibinfo  {journal}
  {Phys. Fluids}\ }\textbf {\bibinfo {volume} {24}},\ \bibinfo {pages} {102107}
  (\bibinfo {year} {2012})},\ \bibinfo {note} {note that a term was missed in
  the variation of $F$ and a correction is contained in the appendix of
  \cite{thap2016prf}.}\BibitemShut {Stop}%
\bibitem [{\citenamefont {Thiele}\ \emph {et~al.}(2016)\citenamefont {Thiele},
  \citenamefont {Archer},\ and\ \citenamefont {Pismen}}]{thap2016prf}%
  \BibitemOpen
  \bibfield  {author} {\bibinfo {author} {\bibfnamefont {U.}~\bibnamefont
  {Thiele}}, \bibinfo {author} {\bibfnamefont {A.~J.}\ \bibnamefont {Archer}},
  \ and\ \bibinfo {author} {\bibfnamefont {L.~M.}\ \bibnamefont {Pismen}},\
  }\href {\doibase 10.1103/PhysRevFluids.1.083903} {\bibfield  {journal}
  {\bibinfo  {journal} {Phys. Rev. Fluids}\ }\textbf {\bibinfo {volume} {1}},\
  \bibinfo {pages} {083903} (\bibinfo {year} {2016})}\BibitemShut {NoStop}%
\bibitem [{\citenamefont {Warner}\ \emph {et~al.}(2002)\citenamefont {Warner},
  \citenamefont {Craster},\ and\ \citenamefont {Matar}}]{WaCM2002pof}%
  \BibitemOpen
  \bibfield  {author} {\bibinfo {author} {\bibfnamefont {M.}~\bibnamefont
  {Warner}}, \bibinfo {author} {\bibfnamefont {R.}~\bibnamefont {Craster}}, \
  and\ \bibinfo {author} {\bibfnamefont {O.}~\bibnamefont {Matar}},\
  }\href@noop {} {\bibfield  {journal} {\bibinfo  {journal} {Phys. Fluids}\
  }\textbf {\bibinfo {volume} {14}},\ \bibinfo {pages} {4040} (\bibinfo {year}
  {2002})}\BibitemShut {NoStop}%
\bibitem [{\citenamefont {Jensen}\ and\ \citenamefont
  {Grotberg}(1992)}]{JeGr1992jfm}%
  \BibitemOpen
  \bibfield  {author} {\bibinfo {author} {\bibfnamefont {O.}~\bibnamefont
  {Jensen}}\ and\ \bibinfo {author} {\bibfnamefont {J.}~\bibnamefont
  {Grotberg}},\ }\href@noop {} {\bibfield  {journal} {\bibinfo  {journal} {J.
  Fluid Mech.}\ }\textbf {\bibinfo {volume} {240}},\ \bibinfo {pages} {259}
  (\bibinfo {year} {1992})}\BibitemShut {NoStop}%
\bibitem [{\citenamefont {Sharma}(1993)}]{Shar1993l}%
  \BibitemOpen
  \bibfield  {author} {\bibinfo {author} {\bibfnamefont {A.}~\bibnamefont
  {Sharma}},\ }\href@noop {} {\bibfield  {journal} {\bibinfo  {journal}
  {Langmuir}\ }\textbf {\bibinfo {volume} {9}},\ \bibinfo {pages} {3580}
  (\bibinfo {year} {1993})}\BibitemShut {NoStop}%
\bibitem [{\citenamefont {Brochard-Wyart}\ \emph {et~al.}(1991)\citenamefont
  {Brochard-Wyart}, \citenamefont {di~Meglio}, \citenamefont {Qu{Že}r{Že}},\
  and\ \citenamefont {de~Gennes}}]{BMQ+1991l}%
  \BibitemOpen
  \bibfield  {author} {\bibinfo {author} {\bibfnamefont {F.}~\bibnamefont
  {Brochard-Wyart}}, \bibinfo {author} {\bibfnamefont {J.-M.}\ \bibnamefont
  {di~Meglio}}, \bibinfo {author} {\bibfnamefont {D.}~\bibnamefont
  {Qu{Že}r{Že}}}, \ and\ \bibinfo {author} {\bibfnamefont {P.-G.}\ \bibnamefont
  {de~Gennes}},\ }\href@noop {} {\bibfield  {journal} {\bibinfo  {journal}
  {Langmuir}\ }\textbf {\bibinfo {volume} {7}},\ \bibinfo {pages} {335 }
  (\bibinfo {year} {1991})}\BibitemShut {NoStop}%
\bibitem [{\citenamefont {Oron}\ \emph {et~al.}(1997)\citenamefont {Oron},
  \citenamefont {Davis},\ and\ \citenamefont {Bankoff}}]{OrDB1997rmp}%
  \BibitemOpen
  \bibfield  {author} {\bibinfo {author} {\bibfnamefont {A.}~\bibnamefont
  {Oron}}, \bibinfo {author} {\bibfnamefont {S.}~\bibnamefont {Davis}}, \ and\
  \bibinfo {author} {\bibfnamefont {S.}~\bibnamefont {Bankoff}},\ }\href
  {\doibase 10.1103/RevModPhys.69.931} {\bibfield  {journal} {\bibinfo
  {journal} {Rev. Mod. Phys.}\ }\textbf {\bibinfo {volume} {69}},\ \bibinfo
  {pages} {931} (\bibinfo {year} {1997})}\BibitemShut {NoStop}%
\bibitem [{\citenamefont {Thiele}(2010)}]{Thie2010jpcm}%
  \BibitemOpen
  \bibfield  {author} {\bibinfo {author} {\bibfnamefont {U.}~\bibnamefont
  {Thiele}},\ }\href {\doibase 10.1088/0953-8984/22/8/084019} {\bibfield
  {journal} {\bibinfo  {journal} {J. Phys.: Condens. Matter}\ }\textbf
  {\bibinfo {volume} {22}},\ \bibinfo {pages} {084019} (\bibinfo {year}
  {2010})}\BibitemShut {NoStop}%
\bibitem [{\citenamefont {Yin}\ \emph {et~al.}(2017)\citenamefont {Yin},
  \citenamefont {Sibley}, \citenamefont {Thiele},\ and\ \citenamefont
  {Archer}}]{YSTA2017pre}%
  \BibitemOpen
  \bibfield  {author} {\bibinfo {author} {\bibfnamefont {H.}~\bibnamefont
  {Yin}}, \bibinfo {author} {\bibfnamefont {D.}~\bibnamefont {Sibley}},
  \bibinfo {author} {\bibfnamefont {U.}~\bibnamefont {Thiele}}, \ and\ \bibinfo
  {author} {\bibfnamefont {A.}~\bibnamefont {Archer}},\ }\href {\doibase
  10.1103/PhysRevE.95.023104} {\bibfield  {journal} {\bibinfo  {journal} {Phys.
  Rev. E}\ }\textbf {\bibinfo {volume} {95}},\ \bibinfo {pages} {023104}
  (\bibinfo {year} {2017})}\BibitemShut {NoStop}%
\bibitem [{\citenamefont {Snoeijer}\ and\ \citenamefont
  {Andreotti}(2008)}]{SA2008pof}%
  \BibitemOpen
  \bibfield  {author} {\bibinfo {author} {\bibfnamefont {J.~H.}\ \bibnamefont
  {Snoeijer}}\ and\ \bibinfo {author} {\bibfnamefont {B.}~\bibnamefont
  {Andreotti}},\ }\href@noop {} {\bibfield  {journal} {\bibinfo  {journal}
  {Physics of Fluids}\ }\textbf {\bibinfo {volume} {20}},\ \bibinfo {pages}
  {057101} (\bibinfo {year} {2008})}\BibitemShut {NoStop}%
\bibitem [{\citenamefont {Bormashenko}(2009)}]{Borm2009csaea}%
  \BibitemOpen
  \bibfield  {author} {\bibinfo {author} {\bibfnamefont {E.}~\bibnamefont
  {Bormashenko}},\ }\href {\doibase 10.1016/j.colsurfa.2009.04.054} {\bibfield
  {journal} {\bibinfo  {journal} {Colloid Surf. A-Physicochem. Eng. Asp.}\
  }\textbf {\bibinfo {volume} {345}},\ \bibinfo {pages} {163} (\bibinfo {year}
  {2009})}\BibitemShut {NoStop}%
\bibitem [{\citenamefont {Dietrich}(1988)}]{Diet1988}%
  \BibitemOpen
  \bibfield  {author} {\bibinfo {author} {\bibfnamefont {S.}~\bibnamefont
  {Dietrich}}\ }(\bibinfo  {publisher} {Academic Press},\ \bibinfo {address}
  {London},\ \bibinfo {year} {1988})\ p.~\bibinfo {pages} {1}\BibitemShut
  {NoStop}%
\bibitem [{\citenamefont {Schick}(1990)}]{Schi1990}%
  \BibitemOpen
  \bibfield  {author} {\bibinfo {author} {\bibfnamefont {M.}~\bibnamefont
  {Schick}},\ }\enquote {\bibinfo {title} {Introduction to wetting
  phenomena},}\ \ (\bibinfo  {publisher} {Elsevier Science Publishers},\
  \bibinfo {address} {North--Holland},\ \bibinfo {year} {1990})\ p.\ \bibinfo
  {pages} {415}\BibitemShut {NoStop}%
\bibitem [{\citenamefont {Tretyakov}\ \emph {et~al.}(2013)\citenamefont
  {Tretyakov}, \citenamefont {M\"{u}ller}, \citenamefont {Todorova},\ and\
  \citenamefont {Thiele}}]{TMTT2013jcp}%
  \BibitemOpen
  \bibfield  {author} {\bibinfo {author} {\bibfnamefont {N.}~\bibnamefont
  {Tretyakov}}, \bibinfo {author} {\bibfnamefont {M.}~\bibnamefont
  {M\"{u}ller}}, \bibinfo {author} {\bibfnamefont {D.}~\bibnamefont
  {Todorova}}, \ and\ \bibinfo {author} {\bibfnamefont {U.}~\bibnamefont
  {Thiele}},\ }\href {\doibase 10.1063/1.4790581} {\bibfield  {journal}
  {\bibinfo  {journal} {J. Chem. Phys.}\ }\textbf {\bibinfo {volume} {138}},\
  \bibinfo {pages} {064905} (\bibinfo {year} {2013})}\BibitemShut {NoStop}%
\bibitem [{\citenamefont {Hughes}\ \emph {et~al.}(2017)\citenamefont {Hughes},
  \citenamefont {Thiele},\ and\ \citenamefont {Archer}}]{HuTA2017jcp}%
  \BibitemOpen
  \bibfield  {author} {\bibinfo {author} {\bibfnamefont {A.~P.}\ \bibnamefont
  {Hughes}}, \bibinfo {author} {\bibfnamefont {U.}~\bibnamefont {Thiele}}, \
  and\ \bibinfo {author} {\bibfnamefont {A.~J.}\ \bibnamefont {Archer}},\
  }\href {\doibase 10.1063/1.4974832} {\bibfield  {journal} {\bibinfo
  {journal} {J. Chem. Phys.}\ }\textbf {\bibinfo {volume} {146}},\ \bibinfo
  {pages} {064705} (\bibinfo {year} {2017})}\BibitemShut {NoStop}%
\bibitem [{\citenamefont {de~Gennes}(1985)}]{Genn1985rmp}%
  \BibitemOpen
  \bibfield  {author} {\bibinfo {author} {\bibfnamefont {P.-G.}\ \bibnamefont
  {de~Gennes}},\ }\href {\doibase 10.1103/RevModPhys.57.827} {\bibfield
  {journal} {\bibinfo  {journal} {Rev. Mod. Phys.}\ }\textbf {\bibinfo {volume}
  {57}},\ \bibinfo {pages} {827} (\bibinfo {year} {1985})}\BibitemShut
  {NoStop}%
\bibitem [{\citenamefont {MacDowell}(2011)}]{MacD2011epjst}%
  \BibitemOpen
  \bibfield  {author} {\bibinfo {author} {\bibfnamefont {L.}~\bibnamefont
  {MacDowell}},\ }\href {\doibase 10.1140/epjst/e2011-01447-6} {\bibfield
  {journal} {\bibinfo  {journal} {Eur. Phys. J. Special Topics}\ }\textbf
  {\bibinfo {volume} {197}},\ \bibinfo {pages} {131} (\bibinfo {year}
  {2011})}\BibitemShut {NoStop}%
\bibitem [{\citenamefont {Hughes}\ \emph {et~al.}(2015)\citenamefont {Hughes},
  \citenamefont {Thiele},\ and\ \citenamefont {Archer}}]{HuTA2015jcp}%
  \BibitemOpen
  \bibfield  {author} {\bibinfo {author} {\bibfnamefont {A.~P.}\ \bibnamefont
  {Hughes}}, \bibinfo {author} {\bibfnamefont {U.}~\bibnamefont {Thiele}}, \
  and\ \bibinfo {author} {\bibfnamefont {A.~J.}\ \bibnamefont {Archer}},\
  }\href {\doibase 10.1063/1.4907732} {\bibfield  {journal} {\bibinfo
  {journal} {J. Chem. Phys.}\ }\textbf {\bibinfo {volume} {142}},\ \bibinfo
  {pages} {074702} (\bibinfo {year} {2015})}\BibitemShut {NoStop}%
\bibitem [{\citenamefont {Bonn}\ \emph {et~al.}(2009)\citenamefont {Bonn},
  \citenamefont {Eggers}, \citenamefont {Indekeu}, \citenamefont {Meunier},\
  and\ \citenamefont {Rolley}}]{BEIM2009rmp}%
  \BibitemOpen
  \bibfield  {author} {\bibinfo {author} {\bibfnamefont {D.}~\bibnamefont
  {Bonn}}, \bibinfo {author} {\bibfnamefont {J.}~\bibnamefont {Eggers}},
  \bibinfo {author} {\bibfnamefont {J.}~\bibnamefont {Indekeu}}, \bibinfo
  {author} {\bibfnamefont {J.}~\bibnamefont {Meunier}}, \ and\ \bibinfo
  {author} {\bibfnamefont {E.}~\bibnamefont {Rolley}},\ }\href {\doibase
  10.1103/RevModPhys.81.739} {\bibfield  {journal} {\bibinfo  {journal} {Rev.
  Mod. Phys.}\ }\textbf {\bibinfo {volume} {81}},\ \bibinfo {pages} {739}
  (\bibinfo {year} {2009})}\BibitemShut {NoStop}%
\bibitem [{\citenamefont {Bastian}\ \emph
  {et~al.}(2008{\natexlab{a}})\citenamefont {Bastian}, \citenamefont {Blatt},
  \citenamefont {Dedner}, \citenamefont {Engwer}, \citenamefont {Kl{\"o}fkorn},
  \citenamefont {Kornhuber}, \citenamefont {Ohlberger},\ and\ \citenamefont
  {Sander}}]{BBD+2008C1}%
  \BibitemOpen
  \bibfield  {author} {\bibinfo {author} {\bibfnamefont {P.}~\bibnamefont
  {Bastian}}, \bibinfo {author} {\bibfnamefont {M.}~\bibnamefont {Blatt}},
  \bibinfo {author} {\bibfnamefont {A.}~\bibnamefont {Dedner}}, \bibinfo
  {author} {\bibfnamefont {C.}~\bibnamefont {Engwer}}, \bibinfo {author}
  {\bibfnamefont {R.}~\bibnamefont {Kl{\"o}fkorn}}, \bibinfo {author}
  {\bibfnamefont {R.}~\bibnamefont {Kornhuber}}, \bibinfo {author}
  {\bibfnamefont {M.}~\bibnamefont {Ohlberger}}, \ and\ \bibinfo {author}
  {\bibfnamefont {O.}~\bibnamefont {Sander}},\ }\href@noop {} {\bibfield
  {journal} {\bibinfo  {journal} {Computing}\ }\textbf {\bibinfo {volume}
  {82}},\ \bibinfo {pages} {103} (\bibinfo {year}
  {2008}{\natexlab{a}})}\BibitemShut {NoStop}%
\bibitem [{\citenamefont {Bastian}\ \emph
  {et~al.}(2008{\natexlab{b}})\citenamefont {Bastian}, \citenamefont {Blatt},
  \citenamefont {Dedner}, \citenamefont {Engwer}, \citenamefont {Kl{\"o}fkorn},
  \citenamefont {Kornhuber}, \citenamefont {Ohlberger},\ and\ \citenamefont
  {Sander}}]{BBD+2008C2}%
  \BibitemOpen
  \bibfield  {author} {\bibinfo {author} {\bibfnamefont {P.}~\bibnamefont
  {Bastian}}, \bibinfo {author} {\bibfnamefont {M.}~\bibnamefont {Blatt}},
  \bibinfo {author} {\bibfnamefont {A.}~\bibnamefont {Dedner}}, \bibinfo
  {author} {\bibfnamefont {C.}~\bibnamefont {Engwer}}, \bibinfo {author}
  {\bibfnamefont {R.}~\bibnamefont {Kl{\"o}fkorn}}, \bibinfo {author}
  {\bibfnamefont {R.}~\bibnamefont {Kornhuber}}, \bibinfo {author}
  {\bibfnamefont {M.}~\bibnamefont {Ohlberger}}, \ and\ \bibinfo {author}
  {\bibfnamefont {O.}~\bibnamefont {Sander}},\ }\href@noop {} {\bibfield
  {journal} {\bibinfo  {journal} {Computing}\ }\textbf {\bibinfo {volume}
  {82}},\ \bibinfo {pages} {121} (\bibinfo {year}
  {2008}{\natexlab{b}})}\BibitemShut {NoStop}%
\bibitem [{\citenamefont {Dijkstra}\ \emph {et~al.}(2014)\citenamefont
  {Dijkstra}, \citenamefont {Wubs}, \citenamefont {Cliffe}, \citenamefont
  {Doedel}, \citenamefont {Dragomirescu}, \citenamefont {Eckhardt},
  \citenamefont {Gelfgat}, \citenamefont {Hazel}, \citenamefont {Lucarini},
  \citenamefont {Salinger}, \citenamefont {Phipps}, \citenamefont
  {Sanchez-Umbria}, \citenamefont {Schuttelaars}, \citenamefont {Tuckerman},\
  and\ \citenamefont {Thiele}}]{DWC+2014ccp}%
  \BibitemOpen
  \bibfield  {author} {\bibinfo {author} {\bibfnamefont {H.~A.}\ \bibnamefont
  {Dijkstra}}, \bibinfo {author} {\bibfnamefont {F.~W.}\ \bibnamefont {Wubs}},
  \bibinfo {author} {\bibfnamefont {A.~K.}\ \bibnamefont {Cliffe}}, \bibinfo
  {author} {\bibfnamefont {E.}~\bibnamefont {Doedel}}, \bibinfo {author}
  {\bibfnamefont {I.~F.}\ \bibnamefont {Dragomirescu}}, \bibinfo {author}
  {\bibfnamefont {B.}~\bibnamefont {Eckhardt}}, \bibinfo {author}
  {\bibfnamefont {A.~Y.}\ \bibnamefont {Gelfgat}}, \bibinfo {author}
  {\bibfnamefont {A.}~\bibnamefont {Hazel}}, \bibinfo {author} {\bibfnamefont
  {V.}~\bibnamefont {Lucarini}}, \bibinfo {author} {\bibfnamefont {A.~G.}\
  \bibnamefont {Salinger}}, \bibinfo {author} {\bibfnamefont {E.~T.}\
  \bibnamefont {Phipps}}, \bibinfo {author} {\bibfnamefont {J.}~\bibnamefont
  {Sanchez-Umbria}}, \bibinfo {author} {\bibfnamefont {H.}~\bibnamefont
  {Schuttelaars}}, \bibinfo {author} {\bibfnamefont {L.~S.}\ \bibnamefont
  {Tuckerman}}, \ and\ \bibinfo {author} {\bibfnamefont {U.}~\bibnamefont
  {Thiele}},\ }\href {\doibase 10.4208/cicp.240912.180613a} {\bibfield
  {journal} {\bibinfo  {journal} {Commun. Comput. Phys.}\ }\textbf {\bibinfo
  {volume} {15}},\ \bibinfo {pages} {1} (\bibinfo {year} {2014})}\BibitemShut
  {NoStop}%
\bibitem [{\citenamefont {Doedel}\ and\ \citenamefont
  {Oldeman}(2009)}]{DOC+2007}%
  \BibitemOpen
  \bibfield  {author} {\bibinfo {author} {\bibfnamefont {E.~J.}\ \bibnamefont
  {Doedel}}\ and\ \bibinfo {author} {\bibfnamefont {B.~E.}\ \bibnamefont
  {Oldeman}},\ }\href@noop {} {\emph {\bibinfo {title} {AUTO07p: Continuation
  and Bifurcation Software for Ordinary Differential Equations}}},\ \bibinfo
  {organization} {Concordia University},\ \bibinfo {address} {Montreal}
  (\bibinfo {year} {2009})\BibitemShut {NoStop}%
\bibitem [{\citenamefont {Frank}\ and\ \citenamefont
  {Garoff}(1995)}]{FrGa1995l}%
  \BibitemOpen
  \bibfield  {author} {\bibinfo {author} {\bibfnamefont {B.}~\bibnamefont
  {Frank}}\ and\ \bibinfo {author} {\bibfnamefont {S.}~\bibnamefont {Garoff}},\
  }\href@noop {} {\bibfield  {journal} {\bibinfo  {journal} {Langmuir}\
  }\textbf {\bibinfo {volume} {11}},\ \bibinfo {pages} {87} (\bibinfo {year}
  {1995})}\BibitemShut {NoStop}%
\bibitem [{\citenamefont {Bardon}\ \emph {et~al.}(1996)\citenamefont {Bardon},
  \citenamefont {Cachile}, \citenamefont {Cazabat}, \citenamefont {Fanton},
  \citenamefont {Valignat},\ and\ \citenamefont {Villette}}]{BCCF1996fd}%
  \BibitemOpen
  \bibfield  {author} {\bibinfo {author} {\bibfnamefont {S.}~\bibnamefont
  {Bardon}}, \bibinfo {author} {\bibfnamefont {M.}~\bibnamefont {Cachile}},
  \bibinfo {author} {\bibfnamefont {A.~M.}\ \bibnamefont {Cazabat}}, \bibinfo
  {author} {\bibfnamefont {X.}~\bibnamefont {Fanton}}, \bibinfo {author}
  {\bibfnamefont {M.~P.}\ \bibnamefont {Valignat}}, \ and\ \bibinfo {author}
  {\bibfnamefont {S.}~\bibnamefont {Villette}},\ }\href {\doibase
  10.1039/FD9960400307} {\bibfield  {journal} {\bibinfo  {journal} {Faraday
  Discuss.}\ ,\ \bibinfo {pages} {307}} (\bibinfo {year} {1996})}\BibitemShut
  {NoStop}%
\bibitem [{\citenamefont {Stoebe}(1996)}]{Stoe1996l}%
  \BibitemOpen
  \bibfield  {author} {\bibinfo {author} {\bibfnamefont {T.~e.~a.}\
  \bibnamefont {Stoebe}},\ }\href {\doibase 10.1021/la950513x} {\bibfield
  {journal} {\bibinfo  {journal} {Langmuir}\ }\textbf {\bibinfo {volume}
  {12}},\ \bibinfo {pages} {337} (\bibinfo {year} {1996})}\BibitemShut
  {NoStop}%
\bibitem [{\citenamefont {Starov}\ \emph {et~al.}(2000)\citenamefont {Starov},
  \citenamefont {Kosvintsev},\ and\ \citenamefont {Velarde}}]{StKV2000jcis}%
  \BibitemOpen
  \bibfield  {author} {\bibinfo {author} {\bibfnamefont {V.~M.}\ \bibnamefont
  {Starov}}, \bibinfo {author} {\bibfnamefont {S.~R.}\ \bibnamefont
  {Kosvintsev}}, \ and\ \bibinfo {author} {\bibfnamefont {M.~G.}\ \bibnamefont
  {Velarde}},\ }\href {\doibase 10.1006/jcis.2000.6851} {\bibfield  {journal}
  {\bibinfo  {journal} {J. Colloid Interface Sci.}\ }\textbf {\bibinfo {volume}
  {227}},\ \bibinfo {pages} {185} (\bibinfo {year} {2000})}\BibitemShut
  {NoStop}%
\bibitem [{\citenamefont {Sharma}\ \emph {et~al.}(2012)\citenamefont {Sharma},
  \citenamefont {Kalita}, \citenamefont {Swanson}, \citenamefont {Corcoran},
  \citenamefont {Garoff}, \citenamefont {Przybycien},\ and\ \citenamefont
  {Tilton}}]{SKSC2012l}%
  \BibitemOpen
  \bibfield  {author} {\bibinfo {author} {\bibfnamefont {R.}~\bibnamefont
  {Sharma}}, \bibinfo {author} {\bibfnamefont {R.}~\bibnamefont {Kalita}},
  \bibinfo {author} {\bibfnamefont {E.}~\bibnamefont {Swanson}}, \bibinfo
  {author} {\bibfnamefont {T.}~\bibnamefont {Corcoran}}, \bibinfo {author}
  {\bibfnamefont {S.}~\bibnamefont {Garoff}}, \bibinfo {author} {\bibfnamefont
  {T.}~\bibnamefont {Przybycien}}, \ and\ \bibinfo {author} {\bibfnamefont
  {R.}~\bibnamefont {Tilton}},\ }\href {\doibase 10.1021/la303639w} {\bibfield
  {journal} {\bibinfo  {journal} {Langmuir}\ }\textbf {\bibinfo {volume}
  {28}},\ \bibinfo {pages} {15212} (\bibinfo {year} {2012})}\BibitemShut
  {NoStop}%
\bibitem [{\citenamefont {Riegler}\ and\ \citenamefont
  {Spratte}(1992)}]{RiSp1992tsf}%
  \BibitemOpen
  \bibfield  {author} {\bibinfo {author} {\bibfnamefont {H.}~\bibnamefont
  {Riegler}}\ and\ \bibinfo {author} {\bibfnamefont {K.}~\bibnamefont
  {Spratte}},\ }\href {\doibase 10.1016/0040-6090(92)90153-3} {\bibfield
  {journal} {\bibinfo  {journal} {Thin Solid Films}\ }\textbf {\bibinfo
  {volume} {210}},\ \bibinfo {pages} {9} (\bibinfo {year} {1992})}\BibitemShut
  {NoStop}%
\bibitem [{\citenamefont {K{\"o}pf}\ \emph {et~al.}(2012)\citenamefont
  {K{\"o}pf}, \citenamefont {Gurevich}, \citenamefont {Friedrich},\ and\
  \citenamefont {Thiele}}]{KGFT2012njp}%
  \BibitemOpen
  \bibfield  {author} {\bibinfo {author} {\bibfnamefont {M.~H.}\ \bibnamefont
  {K{\"o}pf}}, \bibinfo {author} {\bibfnamefont {S.~V.}\ \bibnamefont
  {Gurevich}}, \bibinfo {author} {\bibfnamefont {R.}~\bibnamefont {Friedrich}},
  \ and\ \bibinfo {author} {\bibfnamefont {U.}~\bibnamefont {Thiele}},\ }\href
  {\doibase 10.1088/1367-2630/14/2/023016} {\bibfield  {journal} {\bibinfo
  {journal} {New J. Phys.}\ }\textbf {\bibinfo {volume} {14}},\ \bibinfo
  {pages} {023016} (\bibinfo {year} {2012})}\BibitemShut {NoStop}%
\end{thebibliography}
\end{document}